# HIGHEST WEIGHT UNITARY MODULES FOR NON-COMPACT GROUPS AND APPLICATIONS TO PHYSICAL PROBLEMS.

Juan García-Escudero[a] and Miguel Lorente
*Departamento de Física, Universidad de Oviedo*
33007 Oviedo, Spain.

## I. INTRODUCTION

The study of unitarization of representations for non compact real forms of simple Lie Algebras has been achieved in the past decade by Jakobsen (JA81, JA83) and by Enright, Howe and Wallach (EH83) following different paths but arriving at the same final results.

In order to discuss unitarity we need to introduce a scalar product. This is done in sections II and III introducing a sesquilinear form on the universal enveloping algebra (GL90). Such a sesquilinear form was introduced by Harish-Chandra (HC55), Gel'fand and Kirillov (GK69) and Shapovalov (SH72).

The new developments given to Jakobsen method in GEL90 are contained in sections IV and V. In section VI we summarize the principal results due to Enright Howe and Wallach (EHW method). In section VII we give the possible places for unitarity including those for which the reduction level can't be higher than one and that were not considered in GEL90. We see also in this section and in a explicit way how the Jakobsen method and the EHW method give the same final results.

This type of representations have found applications in physics for a long time (see GK75, 82; GL83, 89; LO86, 89 and references contained therein)

In sections VIII and IX we consider the conformal and de Sitter algebras. In particular the construction of wave equations for conformal multispinors and the wave equations in de Sitter space are reviewed in sections X and XI.

In the Appendix we give the representations of the classical algebras in the subspace $\Omega$ - extending the results obtained in LG84 for the algebra $A_l$.

## II. PRELIMINARIES

Let $g$ be a semisimple Lie algebra over $\mathcal{R}$ and $g^{\mathcal{C}}$ its complexification. Let $B(X,Y) = \text{tr}(\text{ad}X \, \text{ad}Y)$ , $X,Y \in g^{\mathcal{C}}$ be the Killing form. A real form $g_0$ of $g^{\mathcal{C}}$ is called compact if $B(X,X) < 0$ for each $X \in g_0$ and an automorphism $\theta$ of $g^{\mathcal{C}}$ exists such that

$$\theta g_0 \subset g_0 \quad , \quad \theta g \subset g \text{ and } g = k + p \ , \ g_0 = k + ip$$

where $i = \sqrt{-1}$ , $k$ is the set of all $X \in g$ such that $\theta X = X$ and $p$ is the set of all $Y \in g$ such that $\theta Y = -Y$.

Let $k^{\mathcal{C}}$ and $p^{\mathcal{C}}$ be the subspaces of $g^{\mathcal{C}}$ spanned by $k$, $p$ respectively over $\mathcal{C}$. It holds

$$[k^{\mathcal{C}}, k^{\mathcal{C}}] \subset k^{\mathcal{C}} \ , \ [k^{\mathcal{C}}, p^{\mathcal{C}}] \subset p^{\mathcal{C}} \ , \ [p^{\mathcal{C}}, p^{\mathcal{C}}] \subset k^{\mathcal{C}}$$

Let $h$ be a Cartan subalgebra of $g$ and $h^{\mathcal{C}}$ the complexification of $h$. Then $h^{\mathcal{C}}$ is a Cartan subalgebra of $g^{\mathcal{C}}$ and, for the cases considered here (hermitian symmetric spaces of non compact type), holds

$$[h^{\mathcal{C}}, k^{\mathcal{C}}] \subset k^{\mathcal{C}} \quad , \quad [h^{\mathcal{C}}, p^{\mathcal{C}}] \subset p^{\mathcal{C}}$$

For given $g^{\mathcal{C}}, h^{\mathcal{C}}$, let $\Delta$ be the root system of $g^{\mathcal{C}}$ and $\Delta^+$ the system of positive roots. We say that $\alpha$ is compact if $E_\alpha \in k^{\mathcal{C}}$ and non compact if $E_\alpha \in p^{\mathcal{C}}$. The set of compact and non compact roots of $g^{\mathcal{C}}$ with respect to $h^{\mathcal{C}}$ are denoted by $\Delta_c$ and $\Delta_n$ respectively. The set of compact simple roots is denoted by $\Sigma_c$, $\beta$ is the only non compact simple root and $\gamma_r$ is the highest root ( which is a non compact positive root)

Let $k_1 = [k,k]$ and assume that $k$ has a non empty center $\eta$ of dimension one. Then $k = k_1 \oplus \eta$ and $h = (h \leftrightarrow k_1) \oplus \eta$. On the other hand $h^{\mathcal{C}} = (h \leftrightarrow k_1)^{\mathcal{C}} \oplus \eta^{\mathcal{C}}$ is an orthogonal direct sum with respect to the Killing form: for if $H_\mu \in (h \leftrightarrow k_1)^{\mathcal{C}}$ and $H_0 \in \eta^{\mathcal{C}}$

$$(H_\mu, H_0) = ([E_\mu, E_{-\mu}], H_0) = (E_\mu, [E_{-\mu}, H_0]) = 0$$

For $\gamma_1, \gamma_2 \in \Delta$ we use the notation

$$\left\langle \gamma_1, \gamma_2 \right\rangle = \frac{2(\gamma_1, \gamma_2)}{(\gamma_2, \gamma_2)} = \gamma_1\left(H_{\gamma_2}\right)$$

where $(.\,,\,.)$ is the bilinear form on $(h^{\mathcal{C}})^*$ induced by the Killing form on $g^{\mathcal{C}}$.

Let $u(g^{\mathcal{C}})$ be the universal enveloping algebra of $g^{\mathcal{C}}$, $\Lambda \in (h^{\mathcal{C}})^*$ and $R = \frac{1}{2} \sum_{\alpha \in \Delta_+} \alpha$ .

The Verma module $M_\Lambda$ of highest weight $\Lambda$ (VE68) is defined to be $M_\Lambda = u(g^{\mathcal{C}}) \subseteq I_\Lambda$ where $I_\Lambda$ is the left ideal generated by the elements $(H - \Lambda(H))$ , $H \in h^{\mathcal{C}}$ and the set of generators $X\gamma$ with $\gamma \in \Delta^+$. To fix a basis on $(h^{\mathcal{C}})^*$ we choose the set of compact simple roots $\Sigma_c$ for the space $((h \leftrightarrow k_1)^{\mathcal{C}})^*$ and one element $\varepsilon \in (\eta^{\mathcal{C}})^*$ for which

$$\bullet \varepsilon, \mu_i \circledR = 0 \ , \ \forall \mu \in \Sigma_c \text{ and } \bullet \varepsilon, \gamma_r \circledR = 1 \ ,$$

then each $\Lambda \in (h^{\mathcal{C}})^*$ may be written as $\Lambda = \Lambda_0 + \lambda \varepsilon$ , where $\Lambda_0$ satisfies $\bullet \Lambda, \mu_i \circledR = \bullet \Lambda_0, \mu_i \circledR \ \forall \mu_i \in \Sigma_c$. If we choose a normalization for $\Lambda_0$ of the type $\bullet \Lambda_0, \gamma_r \circledR = 0$ , from the last decomposition of $\Lambda$ we conclude that $\bullet \Lambda, \gamma_r \circledR = \lambda$ . The relations $\bullet \Lambda, \mu_i \circledR = \bullet \Lambda_0, \mu_i \circledR$ and $\bullet \Lambda_0, \gamma_r \circledR = 0$ fix $\Lambda_0$ uniquely. In the following we consider $\Lambda_0$ to be $k_1$-dominant and integral, that is $\bullet \Lambda_0, \mu_i \circledR = n_i$ , where $n_i$ are non negative integers.

Now, if $M_\Lambda$ is a Verma module, $L_\Lambda$ an invariant submodule, and $L_\Lambda = M_\Lambda / L_\Lambda$ a quotient module and if $\rho_\Lambda = M_\Lambda$, $L_\Lambda$, $L_\Lambda$ is irreducible then we say that $\rho_\Lambda$ is infinitesimally unitary if there exists a scalar product $(\ ,\ )$ on the carrier space $V$ of $\rho_\Lambda$ such that

$$( u, \rho_\Lambda (X) w ) = - ( \rho_\Lambda (X) u, w )$$

for all $X \in g$ and $u, w \in V$. The above condition is called $g$- invariance.

In a Verma module this scalar product is induced by a sesquilinear form.

## III. SCALAR PRODUCT ON THE ENVELOPING ALGEBRA OF A SEMISIMPLE LIE ALGEBRA

Let $g^\mathcal{C}$ denote a complex semisimple Lie algebra of rank $l$ with diagonal elements $H_i$ ($i = 1, 2, ... l$) for its Cartan subalgebra, shift operators $E_\alpha$ associated to each positive root $\alpha$, shift operators $E_{-\alpha}$ associated to each negative root-$\alpha$. Then the following canonical commutation relations hold:

$$\left[ H_i , E_{\pm\alpha} \right] = \pm \alpha_i\, E_{\pm\alpha}$$

$$\left[ E_\alpha , E_{-\alpha} \right] = \sum_1^l \alpha_i\, H_i \equiv H_\alpha$$

$$\left[ E_\alpha , E_\beta \right] = \begin{cases} N_{\alpha\beta}\, E_{\alpha+\beta} &, \quad \text{if } \alpha + \beta \text{ is a root} \\ 0 &, \qquad\qquad \text{if } \alpha + \beta \text{ is not a root} (\alpha + \beta \neq 0) \end{cases}$$

where $\alpha_i$ denotes the $i$ component of the root $\alpha$.

According to the Poncaré-Birkchoff-Witt theorem (DI74) a basis for the universal enveloping algebra $u(g^\mathcal{C})$ of $g^\mathcal{C}$ can be chosen as the following set of ordered tensor products of the vectors $H_\alpha$, $E_\alpha$, $E_{-\alpha}$ namely,

$$\Omega = \left\{ \mathbf{1} , E_{-\alpha}^m E_{-\beta}^n \dots E_{-\delta}^p\, H_\alpha^q H_\beta^r \dots H_\delta^s\, E_\alpha^t E_\beta^u \dots E_\delta^v \right\}$$

$$\equiv X\,(m, n, \dots, p, q, r, \dots, s, t, u, \dots, v\,)$$

where $m, n, p, q, r, s, t, u, v$, are non negative integers. The symbol $\mathbf{1}$ represents the identity element of the enveloping algebra, i.e. when all exponents are equal to zero.

The basis for the enveloping algebra can be written as

$$\Omega = \Omega_-\, \Omega_0\, \Omega_+$$

where $\Omega_-$ is the enveloping algebra of the vectors $E_{-\alpha}$ associated with negative roots, $\Omega_+$ is the enveloping algebra of the vectors $E_\alpha$ associated with positive roots, and $\Omega_0$ is the enveloping algebra of the elements $H_\alpha$ of the Cartan subalgebra, $h^\mathcal{C}$.

Now for the elements $X$ of the basis $\Omega$ we define the mapping $\gamma : \Omega \to \Omega_0$ by

$$\gamma (X) = \begin{cases} X &, \quad \text{if } X \in \Omega_0 \\ 0 &, \quad \text{otherwise} \end{cases}$$

Let $X$ be a general element of the enveloping algebra $u(g^\mathcal{C})$ expressed in terms of the basis $\Omega$

$$X = \sum c_{m\,n\ \ p\,q\,r\,s\,t\,u\,v}\, X\,(m, n, \dots, p, q, r, \dots, s, t, u, \dots, v)$$

where the $c's$ are complex numbers; then

$$\gamma(X) = X \quad , \quad \text{if} \quad X = \sum c_{o,\ldots,\, q,r,s,o\,\ldots}\, X\,(o,\,\ldots,\,o,q,r,\,\ldots,\,s,o,\,\ldots)$$

$$\gamma(X) = 0 \quad , \quad \text{otherwise}$$

defines a projection of the elements of $u(g^{\,\ell})$ into the subspace expanded by the basis elements of $\Omega_0$.

If we adscribe a weight $\mu$ to the elements of $\Omega$ by the definition

$$\mu = (t-m)\alpha + (u-m)\beta + \ldots + (v-p)\,\delta$$

it is obvious that $\gamma = 0$ if $\mu \neq 0$

Let $\Lambda$ denote a linear form in the dual space of the Cartan subalgebra such that

$$\Lambda\left(H_\alpha\right) = \frac{2\left(\Lambda,\,\alpha\right)}{\left(\alpha,\,\alpha\right)}$$

where $(\,,\,)$ is the scalar product in the dual space $(h^{\,\ell})^*$ induced by the Cartan-Killing form. Therefore $\Lambda$ must be considered as a vector (with complex componentes) in the $l$-dimensional dual space. We will use the composition $\Lambda \cdot \gamma \equiv \xi_\Lambda$.

Given a real semisimple Lie algebra $g$ and its complexification $g^{\,\ell}$, let $\sigma$ denote the conjugation of $g^{\,\ell}$ with respect to $g$:

$$\sigma: g^{\,\ell} \oslash g^{\,\ell},$$

$$\sigma(X+iY) = X - iY \quad \text{for} \quad X,Y \in g$$

Then $-\sigma$ can be extended to an antilinear antiautomorphism $\eta$ of $u(g^{\,\ell})$ as follows:

$$\eta(1) = 1,$$

$$\eta(X) = -\sigma(X),$$

$$\eta(XY \ldots Z) = \eta(Z) \ldots \eta(Y)\,\eta(X).$$

We can define a sesquilinear form $S$ on $u(g^{\,\ell}) \times u(g^{\,\ell})$ for any $\Lambda \in h^*$ as follows:

$$S: u(g^{\,\ell}) \times u(g^{\,\ell}) \oslash C,$$

$$S\,(X,Y) = \xi_\Lambda\{\eta(X)Y\}, \quad X,Y \in u(g^{\,\ell})$$

In general, i.e., on $u(g^{\,\ell})$ and for arbitrary $\Lambda$, $S$ is degenerate and indefinite for a given $g$. However, for some $\Lambda$ it may induce a scalar product on irreducible quotients of Verma modules of highest weight $\Lambda$ and unitarize the finite-dimensional irreducible representations if $g$ is compact (see HC55). On the other hand, if $g$ is noncompact, then for some $\Lambda$ it may induce a scalar product on irreducible Verma modules or on irreducible submodules of Verma modules of highest weight $\Lambda$ and unitarize infinite-dimensional irreducible representations of $g$ (KV78, HC55).

If $V$ is the irreducible quotient of a Verma module, then $\pi_\Lambda$ is infinitesimally unitary with respect to the scalar product induced by $S$ if $\xi_\Lambda\{\eta(z)z\}$ is real and non-negative for every $z \in V$ (see HC55).

## IV.- JAKOBSEN METHOD

In the following we are going to reformulate the Jakobsen method to calculate the modules $M_\Lambda$ that are unitarizable by using a diagramatic representation of $\Delta_n^+$

The modules $M_\Lambda$ are determined by $\Lambda_0$ and $\lambda$ where $\Lambda_0$ is $k_1$-dominant and integral and $\lambda \in \mathcal{R}$

There exists a way to represent the set $\Delta_n^+$ by means of bidimensional diagrams in the following way: one begins with $\beta$ and draws an arrow originating at $\beta$ for each compact simple root $\mu_i$ such that $\beta + \mu_i \in \Delta_n^+$.

*Lemma 4.1* : of JA83 shows that $i \le 2$. We suppose for simplicity that $i = 2$. Then one draws two arrows: one originating at $\beta + \mu_1$ and parallel to $\mu_2$ and another originating at $\beta + \mu_2$ and parallel to $\mu_1$, both arrows point towards $\beta + \mu_1 + \mu_2$ which is also a root. The next step would be to add compact simple roots to the non compact roots previously obtained by keeping those that are non compact roots. Continuing along these lines the diagram may be completed.

For the description of the possible places for unitarity, Jakobsen uses the Bernshtein, Gel'fand and Gel'fand theorem (BG71). This theorem describes the circumstances under which the irreducible quotient $L_\xi$ of a highest weight module can occur in the Jordan-Hölder series JH($M_\Lambda$) of another:

*Definition 1:* Let $\xi$, $\Lambda \in (h^\ell)^*$ . A sequence of roots $\alpha_1,...,\ \alpha_k \in \Delta^+$ is said to satisfy condition (A) for the pair ( $\xi + R$, $\Lambda + R$) if

    a) $\xi + R = \sigma_{\alpha_\kappa} ... \sigma_{\alpha_1} (\Lambda + R)$ where $\sigma_{\alpha_i}$ is the Weyl reflexion with respect to $\alpha_i$

    b) Take $\xi_0 \equiv \Lambda$ , $\xi_i + R = \sigma_{\alpha_i} .. \sigma_{\alpha_1} (\Lambda + R)$

    Then $\xi_{i-1} - \xi_i = n_i \alpha_i$ , $n_i \in N$

*Theorem 2:* (Bernshtein, Gel'fand and Gel'fand); Let $\xi$ , $\Lambda \in (h^\ell)^*$ and let $L_\xi$ , $M_\Lambda$ two Verma modules. Then $L_\xi \in$ JH($M_\Lambda$) if and only if there exists a sequence $\alpha_1,...,\alpha_k \in \Delta^+$ satisfying condition (A) for the pair ( $\xi + R$, $\Lambda + R$)

On the other hand, under some conditions the $\alpha_i$ 's may be considered as non compact ones:

*Proposition 3:* Let $\xi$, $\Lambda \in (h^\ell)^*$ and assume that the sequence $\alpha_1,...,\ \alpha_k$ satisfies condition (A) for the pair ( $\xi + R$, $\Lambda + R$). If $\xi$ is $k_1$-dominant we may assume that $\alpha_i \in \Delta_n^+$ , $i = 1...k$.

Let $V_{\Lambda_0}$ be an irreducible finite-dimensional $u(k_1^\ell)$-module with highest weight $\Lambda_0$. We first consider the $u(k_1^\ell)$-module $p^- \otimes V_{\Lambda_0}$. The highest weights on $p^- \otimes V_{\Lambda_0}$ are of the form $\Lambda_0 - \alpha$ for certain $\alpha \in \Delta_n^+$ which we will describe in terms of the Jakobsen diagrams.

We now describe the method:

**i)** Let $\alpha \in \Delta_n^+$ and assume $\alpha - \mu_j \in \Delta_n^+$ for $\mu_j \in \Sigma_c$ ; $j = 1 ,..., i$ and $i \le 2$.

    Then $\Lambda_0 - \alpha$ is a highest weight for the $u(k_1^\ell)$-module $p^- \otimes V_{\Lambda_0}$ if and only if for all $j = 1,..., i$     $\Lambda_0(H_{\mu_j}) \equiv \langle \Lambda_0, \mu_j \rangle \ge \max \{ 1, \langle \alpha, \mu_j \rangle \}$

    Recall that $\Lambda_0$ is fixed by given integers $\langle \Lambda_0, \mu_i \rangle$, $\mu_i \in \Sigma_c$ and $\langle \Lambda_0, \gamma_r \rangle = 0$

**ii)** For those $\alpha \in \Delta_n^+$ of step i) let $\lambda_\alpha \in R$ be determined by the equation

$$\langle \Lambda + R, \ \alpha \rangle = (\Lambda_0 + \lambda_\alpha \varepsilon + R) (H_\alpha) = 1$$

Let $\lambda_0$ denote the samallest among those $\lambda_\alpha$'s , and let $\alpha_0$ denote the corresponding element of $\Delta_n^+$. We now define the following sets:

$$C_{\alpha_0}^+ = \{ \alpha \in \Delta_n^+ / \alpha \ge \alpha_0 \} \quad \text{and} \quad C_{\alpha_0}^- = \{ \alpha \in \Delta_n^+ / \alpha \le \alpha_0 \}.$$

The way in which those sets appear in the diagram of $\Delta_n^+$ suggest that we can call $C_{\alpha_0}$ and $C_{\alpha_0}$ the forward and backward cone, respectively, at $\alpha_0$.

**iii**) Let $\omega_q = n_1\,\alpha_1 + ... + n_r\,\alpha_r$, $n_i \in N$. If $\alpha_1, ..., \alpha_r \in \Delta_n^+$ satisfies condition (A) for the pair $(\Lambda - \omega_q + R, \Lambda + R)$ where $\Lambda = \Lambda_0 + \lambda_q \varepsilon$ and $\Lambda_0 - \omega_q$ is the weight of a highest weight vector $q$ in the $u(k_1^{\varphi})$-module $u(p^-) \otimes V_{\Lambda_0}$ and $\lambda_q < \lambda_0$, then

$$\alpha_i \in C^+{}_{\alpha_0} \quad \forall\, i = 1...r$$

**iv**) The $\alpha_i$'s appearing in $\omega_q$ must satisfy certain conditions which we now describe. Because inner products between positive noncompact roots are non negative and because $\lambda_q < \lambda_0$, it follows that for $\alpha_i \in \Delta_n^+$

$$\bullet \Lambda_0 + \lambda_0 \varepsilon + R\,,\,\alpha_i \circledR > \bullet \Lambda_0 + \lambda_q \varepsilon + R\,,\,\alpha_i \circledR > 0$$

On the other hand, to check the $k_1$-dominance of $\Lambda_0 - \omega_q$ i.e.

$$\bullet \Lambda_0 - \omega_q\,,\,\mu_j\,\circledR \geq 0 \qquad \forall\,\mu_j \in \Sigma_c$$

is useful to have in mind that if a compact simple root $\mu$ is pointing towards a non compact positive root $\alpha$ in the diagram then $\bullet \alpha\,,\mu\,\circledR > 0$ and if $\mu$ arises outwards $\alpha$ then $\bullet \alpha\,,\mu\,\circledR < 0$ provided that the same $\mu$ both is not pointing towards and arises outwards $\alpha$.

**v**) $M_\Lambda$ with $\Lambda = \Lambda_0 + \lambda_0 \varepsilon$ is unitarizable. The value $\lambda = \lambda_0$ is called the last possible place for unitarity, because for $\lambda > \lambda_0$ there is no unitarity. The description of the general situation follows by forming tensor products of $M_\Lambda$ with the unitary module $M_{\lambda_s \varepsilon}$ corresponding to $\Lambda_0 = 0$. The restriction of $M_\Lambda \otimes M_{\lambda_s \varepsilon}$ to the diagonal is the unitarizable module $M_{\Lambda'}$ with $\Lambda' = \Lambda_0 + (\lambda_0 + \lambda_s)\,\varepsilon$.

This means that if we want to unitarize we must take the quotient space with respect to the invariant subspace generated by the highest weight vector corresponding to $\lambda_0 + \lambda_s$ which is a second order polynomial: this polynomial will be missing.

The modules $M_{\Lambda''}$ with $\Lambda'' = \Lambda_0 + \lambda\varepsilon$, $\lambda_0 + \lambda_s < \lambda < \lambda_0$ are not unitarizable. For $\lambda_0 + 2\lambda_s$ we may have unitarity and there will be a third order missing polynomial, while there is no unitarity for $\lambda_0 + 2\lambda_s < \lambda < \lambda_0 + \lambda_s$.

Continuing along these lines we arrive at the first possible place for non unitarity which corresponds to $\lambda = \lambda_0 + u\lambda_s$ (we call $u + 1$ the reduction level) and all representations with $\lambda < \lambda_0 + u\lambda_s$ are unitary.

The following diagram illustrates the possible places for unitarity (in the next section we will see that $\lambda_s < 0$).

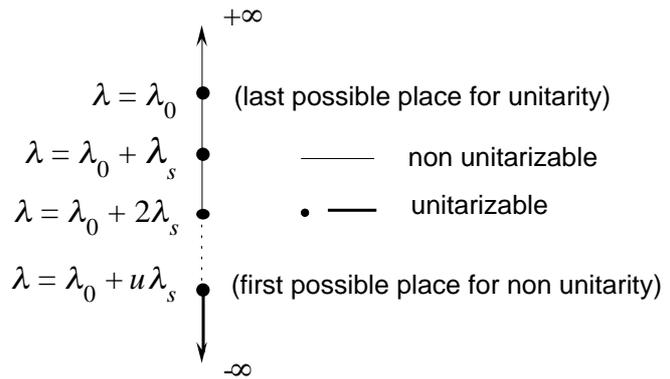

### V.- SOME DEFINITIONS AND NOTATIONS

From the Jakobsen diagrams (see the examples in Section V) we observe that all positive noncompact root can be expresed as

$$\alpha = \beta + \mu_{i_1} + ... + \mu_{i_k}\,,\quad \mu_{i_m} \in \Sigma_c \quad m = 1,...k$$

Thus, given $\alpha$ on this way we define its "height" $H_\alpha$ as $k+1$ (the roots $\mu_{i_m}$ may be repeated)

On the other hand given the decomposition $\Lambda = \Lambda_0 + \lambda\varepsilon$ we may relate the products $\bullet\Lambda, \alpha\circledR$ and $\bullet\Lambda_0, \alpha\circledR$, $\alpha \in \Delta_n^+$· in the following way

In fact

$$\bullet\Lambda, \alpha\circledR = \bullet\Lambda_0, \alpha\circledR + \lambda\,\overline{\phantom{xxx}}\;\frac{}{(\alpha,\alpha)}$$

$$\bullet\Lambda, \alpha\circledR = \bullet\Lambda_0, \alpha\circledR + \lambda\bullet\varepsilon, \alpha\circledR$$

and, decomposing $\alpha = \gamma_r - \underset{\mu_i \in \Sigma_c}{\phantom{xx}}$ (see Jakobsen diagrams) then

$$\langle\varepsilon, \alpha\rangle = \frac{2(\varepsilon, \gamma_r)}{(\alpha, \alpha)} = \frac{(\gamma_r, \gamma_r)}{(\alpha, \alpha)}$$

where we use the fact that $\bullet\varepsilon, \gamma_r\circledR = 1$ and $\bullet\varepsilon, \mu\circledR = 0$ $\forall\mu \in \Sigma_c$

By means of a direct calculation we obtain the following useful expresions for the products $\bullet R, \alpha\circledR$ and $\bullet\Lambda, \alpha\circledR$ that will be needed in next section

**a)** su $(p, q)$, so*(2n), so $(2n-2, 2)$, $e_6$ and $e_7$

$$\bullet R, \alpha\circledR = H_\alpha$$

$$\bullet\Lambda, \alpha\circledR = (\Lambda_0, \alpha) + \lambda$$

**b)** sp $(n, \mathcal{R})$

If $\alpha$ is short $\bullet R, \alpha\circledR = H_\alpha + 1$

$$\bullet\Lambda, \alpha\circledR = (\Lambda_0, \alpha) + 2\lambda$$

If $\alpha$ is long $\bullet R, \alpha\circledR = \frac{1}{2}\{H_\alpha + 1\}$

$$\bullet\Lambda, \alpha\circledR = \frac{1}{2}(\Lambda_0, \alpha) + \lambda$$

**c)** so $(2n-1, 2)$

Let $\alpha_1$ be the short noncompact positive root:

If $\alpha$ is long 
$$\begin{cases} \text{i) } C_\alpha^+ \subset C_{\alpha_1}^+ \begin{cases} \langle R, \alpha\rangle = H_\alpha + 1 \\ \langle\Lambda, \alpha\rangle = (\Lambda_0, \alpha) + \lambda \end{cases} \\[2em] \text{ii) } C_{\alpha_1}^+ \subset C_\alpha^+ \begin{cases} \langle R, \alpha\rangle = H_\alpha \\ \langle\Lambda, \alpha\rangle = (\Lambda_0, \alpha) + \lambda \end{cases} \end{cases}$$

If $\alpha = \alpha_1$ 
$$\begin{cases} \langle R, \alpha\rangle = 2H_\alpha - 1 \\ \langle\Lambda, \alpha\rangle = 2(\Lambda_0, \alpha) + 2\lambda \end{cases}$$

From the Jakobsen diagrams we see that all roots of the same height are in an horizontal line.

In the figure 1 such a roots are inside a little circle and we may localize them by means of a subindex $j$ which is equal to the height of the root and an ordenation superindex $i$ which is equal to one, for the root placed at the right branch of the cone generated by $\alpha_0$, and it increases from unit to unit when we are going toward the left branch. In this way we will write $\alpha_j^i$.

In order to calculate the parameter $\hat\lambda_s$ in step **v**) we

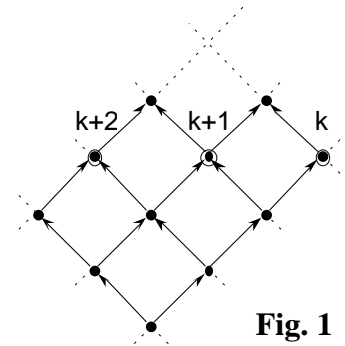

**Fig. 1**

make use of the following

*Definition 4* : (Harish-Chandra) : Let $\gamma_1$ be the smallest element of $\Delta_n^+$ and, inductively, let $\gamma_k$ be the smallest element of $\Delta_n^+$ that is orthogonal to $\gamma_1, \ldots, \gamma_{k-1}$. Let $\gamma_1, \ldots, \gamma_t$ be the maximal collection obtained. Then $t$ is the split-rank of $g$.

With our notation $\gamma_1 \equiv \beta$. We use the Jakobsen diagrams to obtain the split rank.

**su(p,q)**

The collection $\gamma_1, \ldots, \gamma_t$ follows by drawing a line from $\beta$ as is indicated in figure 2.

The roots founded are those which are on the line:

$p \leq q$ : $e_p - e_{p+1}$ , $e_{p-1} - e_{p+2}$ ,....., $e_1 - e_{2p}$
$p \geq q$ : $e_p - e_{p+1}$ , $e_{p-1} - e_{p+2}$ ,....., $e_{p-(q-1)} - e_{p+q}$

So, if $p \leq q$ the split rank is $p$ and if $p \geq q$ the split rank is $q$ , therefore

Split rank of $\quad$ su(p, q) = min {p, q}

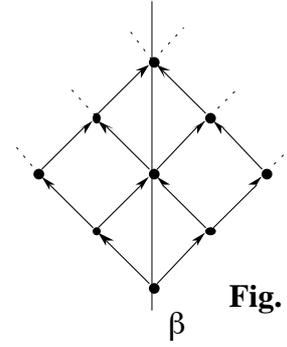

**Fig. 2**

$\beta$

**sp(n,$\mathcal{R}$)**

In the same way as in su (p, q) the collection obtained here is

$$2e_n , 2e_{n-1} ,...., 2e_1$$

Thus

$$\text{Split rank } \text{sp(n,}\mathcal{R}) = n$$

**so*(2n)**

The collection is, in this case

$$e_{n-1} + e_n ,...,e_1 + e_2, \quad \text{if } n \text{ is even}$$
$$e_{n-1} + e_n ,...,e_2 + e_3, \quad \text{if } n \text{ is odd}$$

Then the split rank is $\frac{n}{2}$ if $n$ is even and $\frac{n-1}{2}$ if $n$ is odd:

$$\text{Split rank so*(2n)} = \left[ \frac{n}{2} \right]$$

where [$x$] denotes the largest integer $\leq x$

**so(2n-1, 2) , so(2n-2, 2)**

The split rank is, in both cases, equal to two. The collection is, in this case $e_1 - e_2$ , $e_1 + e_2$

**e$_6$ , e$_7$**

The collection obtained is now

e$_6$ : { $\frac{1}{2}$ ($e_1 - e_2 - e_3 - e_4 - e_5 - e_6 - e_7 + e_8$ ), $\frac{1}{2}$ ($-e_1 + e_2 + e_3 + e_4 - e_5 - e_6 - e_7 + e_8$ )}

then the split rank of $e_6$ is equal to two

e$_7$ : { $e_6 - e_5$ , $e_6 + e_5$ , $e_8 - e_7$ }

thus split rank of $e_7$ is equal to three.

Now let $h^- = \sum_{i=1}^{\notin} \notin H_{\gamma_i}$ and, for $1 \leq j \leq t$ , $t$ being the split rank, let $c_j$ be the number of compact positive roots $\mu$ such that $\mu^{\text{™}} h^- = \frac{1}{2}$ ($\gamma_j - \gamma_i$) , $i < j$. If we consider the most singular non trivial unitary module corresponding to $\Lambda_0 = 0$, then according to Theorem 5.10 in WA79 , $\lambda_q = -\frac{1}{2}c_j$ . A straightforward calculation case by case shows that

$$\lambda_q = (j-1)\,\lambda_s \ , \quad 1 \le j \le t$$

with $\lambda_s$ given in the following table

| | $su(p,q)$ | $sp(n,R)$ | $so^*(2n)$ | $e_6$ | $e_7$ | $so(2n-2,2)$ | $so(2n-1,2)$ |
|---|---|---|---|---|---|---|---|
| $\lambda_s$ | -1 | $-\frac{1}{2}$ | -2 | -3 | -4 | $-n+2$ | $-n+3/2$ |

## VI. EHW METHOD

In this section we summarize the main results given by the EHW method (EP81, EH83).

Those authors use a decomposition of the highest weight in the form $\Lambda = \Pi_0 + z\,\varepsilon$, $z \in \mathcal{R}$, $\langle \Pi_0 + R, \gamma_r \rangle = 0$, $\Pi_0$ $k_1$-dominant and integral, $\varepsilon$ is given in section II (as we see the only difference between this decomposition and the one used by Jakobsen is the normalization choosed for the compact part of the highest weight).

*Theorem 5*: The set of real numbers $z$ such that $M_{\Pi_0 + z\varepsilon}$ is a $g$-module unitarizable, is given in the diagram (see Fig 3).

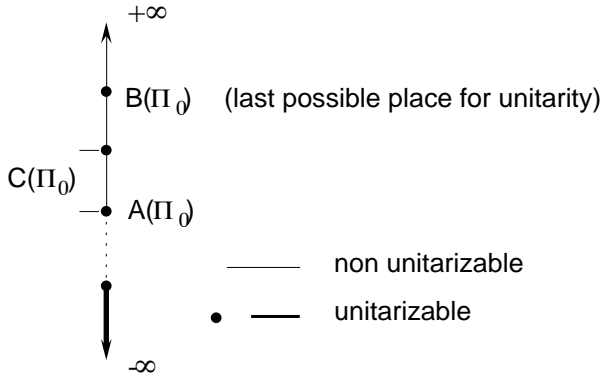

**Fig. 3**

a) The set includes the half line ending at $A(\Pi_0)$.
   The smallest value $z$ with $M_{\Pi_0 + z\varepsilon}$ unitarizable is $z = A(\Pi_0)$, that we call the first reduction point.

b) In addition to the half line there are a number of equally spaced points in the set for which $M_{\Pi_0 + z\varepsilon}$ is unitarizable. The distance between two consecutive points is $C(\Pi_0)$. The set of such a points begins at $A(\Pi_0)$ and end at $B(\Pi_0)$, and we call to its cardinal the reduction level.

In the following we will give the expressions for $A(\Pi_0)$, $B(\Pi_0)$ y $C(\Pi_0)$. To do this we associate root systems to the line $\Pi_0 + z\varepsilon$.

Let $\Delta_c(\Pi_0) = \{ \mu \in \Delta_c \;^{TM}\; \langle \Pi_0, \mu \rangle = 0 \}$ and let $\{ \pm \gamma_r, \Delta_c(\Pi_0) \}$ be the root system of $\Delta$ generated by $\pm \gamma_r$ and $\Delta_c(\Pi_0)$. Decomposing this root system into a disjoint union of simple root systems, let $Q(\Pi_0)$ be the simple root system which contains $\gamma_r$. If $\Delta$ has two lengths and if there exists short compact roots $\mu$ not orthogonal to $Q(\Pi_0)$ with

$\langle \Pi_0, \mu \rangle = 1$ then let $\psi$ be the root system generated by $\pm \gamma_r$, $\Delta_c(\Pi_0)$ and all such $\mu$. Let $T(\Pi_0)$ be the simple component of $\psi$ which contains $\gamma_r$. If $\Delta$ has only one length or if no such $\mu$ exist then $T(\Pi_0) = Q(\Pi_0)$.

We want to see with an example how to construct $Q(\Pi_0)$. We take the Dynkin diagram of $su(p, q)$ and we draw a little circle around the non compact simple root $\beta$:

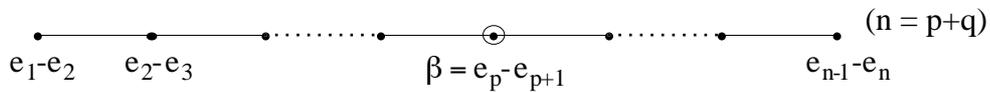

Next we omit the non compact root and we connect $-\gamma_r = e_n - e_1$ by the usual rules (Fig. 4).

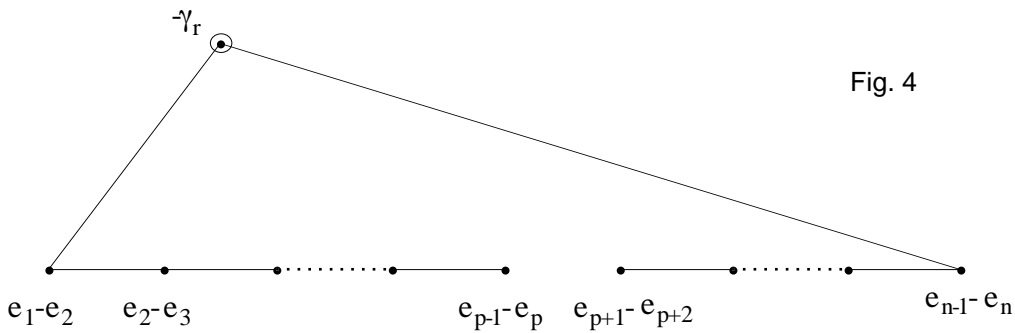

Fig. 4

Now we take the maximal connected diagram which contains $-\gamma_r$ and such that every compact simple root is orthogonal to $\Pi_0$. This subdiagram is the diagram of $Q(\Pi_0)$.

There are only two cases where $Q(\Pi_0) \neq T(\Pi_0)$ and it corresponds to the algebras sp $(n, \mathcal{R})$ and so $(2n - 1, 2)$.

For the calculation of $B(\Pi_0)$ we consider $\Delta_{c,1} = \Delta_c \cap Q(\Pi_0)$ and $\Delta_{c,2} = \Delta_c \cap T(\Pi_0)$ Let $R_{c,1}$ ( respectively $R_{c,2}$ ) be half the sum of roots in $\Delta_{c,1}^+$ (respectively $\Delta_{c,2}^+$ ).

**Theorem 6 :**

    **a)** If $g =$ so $(2n - 1, 2)$ and $Q(\Pi_0) \neq T(\Pi_0)$ then $B(\Pi_0) = 1 + \bullet R_{c,2}, \gamma_r \circledR$

    **b)** In all other cases $B(\Pi_0) = 1 + \bullet R_{c,1} + R_{c,2}, \gamma_r \circledR$.

The constants $C(\Pi_0)$ only depend of $g$ and is equal to $-\lambda_s$, where $\lambda_s$ is the parameter given at the end of section V

The first reduction point $A(\Pi_0)$ is given by:

**Theorem 7 :**

$$A(\Pi_0) = B(\Pi_0) - \Big(\text{split rank of } Q(\Pi_0) - 1\Big).C(\Pi_0)$$

In the following the results for each algebra are given.

**su (p , q)**

We consider $\Lambda = \Pi_0 + z\varepsilon$ with $\Pi_0 = (\Pi_1, \Pi_2, ...., \Pi_n)$, $n = p + q$ ; as $\Pi_0$ is $k_1$-dominant and integral $\Pi_1 \geq \Pi_2 .... \geq \Pi_p$ , $\Pi_{p+1} \geq .... \geq \Pi_n$ and $\Pi_i - \Pi_j \in \mathbb{Z}^+$, $1 \leq i < j \leq p$ or $p+1 \leq i < j \leq n$. On the other hand as $\bullet \Pi_0 + R, \gamma_r \circledR = 0$, $\Pi_1 - \Pi_n + n - 1 = 0$. In addition $\varepsilon = (q/n , q/n , ... , q/n , -p/n , ... , -p/n )$ with $p$ copies of $q/n$ and $q$ of $-p/n$. .For su $(p, q)$ we have:

$$2R_c = \Big(p - 1, p - 3, ... , p - (2p - 1), q - 1, q - 3, ... , q - (2q - 1)\Big)$$

We now suppose that we have the following conditions on the $\Pi_0$ components:

$$\Pi_1 = \Pi_2 = ... = \Pi_i \ \neq \ \Pi_{i+1}$$

$$\Pi_n = \Pi_{n-1} = ... = \Pi_{n-j+1} \ \neq \ \Pi_{n-j}$$

Then the root sistem $Q(\Pi_0)$ has the form :

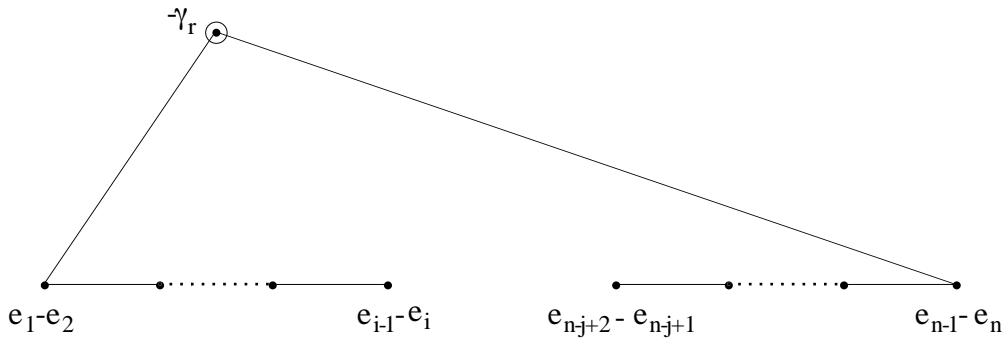

which is of type su $(i, j)$. Then :

$$2R_{c,1} = (i - 1, \ ... , -i + 1, 0, ... , 0, j - 1, ... , -j + 1)$$

In this case we have $T(\Pi_0) = Q(\Pi_0)$, $R_{c,1} + R_{c,2} = 2R_{c,1}$ and $\gamma_r = e_1 - e_n$ ; then applying Theorem 6 :

**Lemma 8 .-** The last possible place for unitarity for su $(p, q)$ is $B(\Pi_0) = i + j - 1$

From the Theorem 7 :  $A(\Pi_0) = i + j - min\,\{i\,,\,j\}$  and the reduction level is

$B(\Pi_0) - A(\Pi_0) + 1$ :

*Lemma 9* .- The first possible place for non unitarity is  $max\,\{i\,,\,j\}$  and the reduction level is

$min\,\{i\,,\,j\}$.

Then we will have :

*Theorem 10*:

$M_{\Pi_0 + z\varepsilon}$  is unitarizable if and only if  $z \le max\,\{i\,,\,j\}$  or  $z$   is an integer and

$\qquad z \le i + j - 1$

## sp(n , $\mathcal{R}$)

In this case   $\Lambda = \Pi_0 + z\varepsilon$ ,  $\Pi_0 = (\Pi_1\,,\,....\,,\,\Pi_n)$    and since it is  $k_1$ -dominant  and integral then  $\Pi_i - \Pi_j \in \mathcal{N}, i < j$ .  From the condition  $\bullet\,\Pi_0 + R\,,\gamma_r \circledR = 0$  we have  $\Pi_1 = -n$ .  On the other hand  $\varepsilon = (1, 1, ... , 1)$.  The sum of positive compact roots is, for  sp(n , $\mathcal{R}$),

$$2R_c = \big(n\,\text{-}1\,,\,n\,\text{-}3\,,\,...\,,\,n\,\text{-}(2n\,\text{-}1)\big)$$

We consider two cases:

***Case I***           $\Pi_1 = \Pi_2 = .... = \Pi_i \ge \Pi_{i+1} + 2$

The root sistem  $Q(\Pi_0)$  is

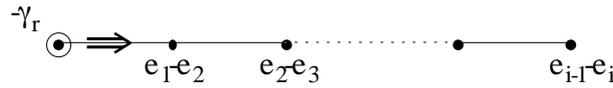

and it is of type  sp($i$ , $\mathcal{R}$), then :    $2R_{c,1} = (i\,\text{-}1\,,\,i\,\text{-}3\,,\,...\,,\,\text{-}\,i+1\,,\,0\,,\,...\,,0\,)$

In this case  $T(\Pi_0) = Q(\Pi_0)$ ,  $R_{c,1} + R_{c,2} = 2R_{c,1}$, and  $\gamma_r = 2e_1$  then, from Theorem 6:

*Lemma 11* .-  The last possible place for unitarity is  $B(\Pi_0) = i$

On the other hand   $A(\Pi_0) = i - (i\,\text{-}1)\,\frac{}{2}$  and the reduction level   $2\big[B(\Pi_0) - A(\Pi_0)\big] + 1$  them from Theorem 7:

*Lemma 12* .- The first possible place for non unitarity is  $\frac{}{2}(i+1)$  and at this point the reduction level is  $i$.

*Theorem 13* .-  $M_{\Pi_0 + z\varepsilon}$  is unitarizable if and only if  $z \le \frac{}{2}(i+1)$   or  $2z \in Z$   with $z \le i$ .

***Case II***           $\Pi_1 = \Pi_2 = .... = \Pi_i$ ,  $\Pi_i - \Pi_{i+1} = 1$ ,  $\Pi_{i+1} = \Pi_{i+2} = ... = \Pi_{i+j} \ne \Pi_{i+j+1}$

The root sistem  $Q(\Pi_0)$  is

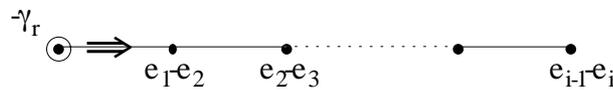

and the root sistem  $T(\Pi_0)$

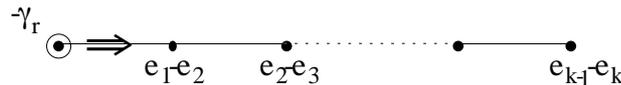

where  $k = i + j$

Then  $Q(\Pi_0)$  is of type   sp($i$ , $\mathcal{R}$)  and  $T(\Pi_0)$  is of type  sp($k$ , $\mathcal{R}$), therefore

$\qquad 2R_{c,1} = (i\,\text{-}1\,,\,i\,\text{-}3\,,\,...\,,\,\text{-}\,i+1\,,\,0\,,\,...\,,\,0)$

$\qquad 2R_{c,2} = (k\,\text{-}1\,,\,k\,\text{-}3\,,\,...\,,\,\text{-}\,k+1\,,\,0\,,\,...\,,\,0)$

From Theorems 6 and 7:

*Lemma 14* .- The last possible place for unitarity is   $B(\Pi_0) = \frac{}{2}(2i + j)$

**Lemma 15** .- The first possible place for non unitarity is $\frac{1}{2}(i + j + 1)$ and the reduction level is $i$ .

**Theorem 16**: $M_{\Pi_0 + z\varepsilon}$ is unitarizable if and only if $z \leq \frac{1}{2}(i + j + 1)$ or $2z \in \mathbb{Z}$ with

$$z \leq \frac{1}{2}(2i + j) .$$

**so\*(2n )**

Given $\Lambda = \Pi_0 + z\varepsilon$ , $\Pi_0 = (\Pi_1 , \dots , \Pi_n )$ and $\varepsilon = \frac{1}{2}( 1, 1, \dots , 1)$ .

As $\Pi_0$ is $k_1$- dominant and integral, $\Pi_i - \Pi_j \in \mathcal{N}$ for $i < j$ . By the normalization •$\Pi_0 + R , \gamma_r \circledR = 0$, we have $\Pi_1 + \Pi_2 = -2n + 3$ . The sum of positive compact roots is :

$$2R_c = \left( n - 1, n - 3, \dots , n - (2n - 1)\right)$$

Now, let the following conditions on the $\Pi_0$ components:

$$\Pi_1 = \Pi_2 = \dots = \Pi_i > \Pi_{i+1} + 1 , \quad i \neq 1$$

Next we consider the root sistem $Q(\Pi_0) = T(\Pi_0)$ which is of type so\*(2i ) , then

$$2R_{c,1} = (i - 1, i - 3, \dots , -i + 1, 0, \dots , 0)$$

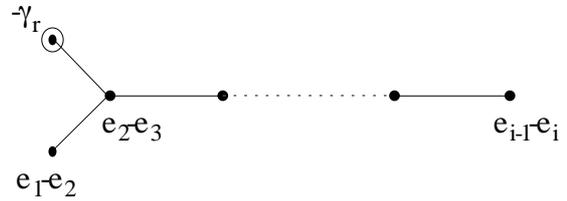

In this case $\gamma_r = e_1 + e_2$ therefore we will have by Theorem 6:

**Lemma 17 :** The last possible place for unitarity is $B(\Pi_0) = 2i - 3$ .

Let $[x]$ denote the biggest integer $\leq$ x. Then

$$A(\Pi_0) = 2i - 3 - 2\left( \left[ \frac{i}{2} \right] - 1 \right)$$

and the reduction level is $\dfrac{\phantom{xx}}{2} + 1$ , then from Theorem 7:

**Lemma 18**:: The first possible place for non unitarity is
$$A(\Pi_0) = \begin{cases} i - 1 & \text{if } i \text{ is even} \\ i & \text{if } i \text{ is odd} \end{cases} \quad \text{and the reduction level } \left[ \frac{i}{2} \right] .$$

**Theorem 19**.: $M_{\Pi_0 + z\varepsilon}$ is unitarizable if and only if $z \leq \begin{cases} i - 1 & \text{if } i \text{ is even} \\ i & \text{if } i \text{ is odd} \end{cases}$ or

$z = 2i - 3 - 2m$ for some integer $m$ with $0 \leq m \leq \left[ \frac{i}{2} \right] - 2$ .

We consider now the case $i = 1$ . So, let $\Pi_1 \neq \Pi_2$ and $j$ the biggest integer for which

$\Pi_2 = \dots = \Pi_{j+1}$ .

The root sistem $Q(\Pi_0) = T(\Pi_0)$ is :

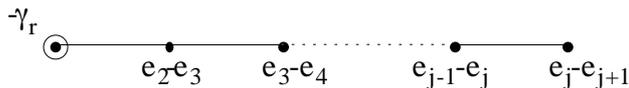

which is of type su $(1, j)$ , then

$$2R_{c,1} = (0, j - 1, j - 3, \dots , -j + 1, 0, \dots , 0)$$

so, applying Theorems 6 and 7 we obtain : $A(\Pi_0) = B(\Pi_0) = j$ then the reduction level is 1 and $M_{\Pi_0 + z\varepsilon}$ is unitarizable if and only if $z \leq j$ .

**so (2n - 1 , 2)**

Let $\Lambda = \Pi_0 + z\varepsilon$ , $\Pi_0 = (\Pi_1 , ... , \Pi_n)$ and $\varepsilon = (1, 0, ... , 0)$ ; $\Pi_0$ is $k_1$ -dominant and integral then $\Pi_2 \geq ... \geq \Pi_n \geq 0$ , $\Pi_i - \Pi_j \in Z$ and $2\Pi_i \in \mathcal{N}$ for all $i$ , $j$ with $2 \leq i , j \leq n$ . From the condition $\bullet \Pi_0 + R$ , $\gamma_r \circledR = 0$ we have $\Pi_1 + \Pi_2 = -2n + 2$ .

We consider two cases

***Case I***       Let the following conditions on the $\Pi_0$ components

$$\Pi_2 = \Pi_3 = ... = \Pi_i \neq \Pi_{i+1} \quad i < n$$

The root sistem $Q(\Pi_0) = T(\Pi_0)$ is the following

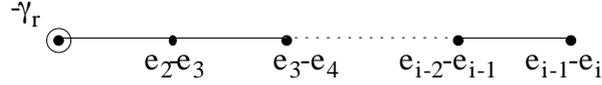

and it is of type su $(1 , i - 1)$ then

$$2R_{c,1} = (0, i - 2 , i - 4 , ... , -i + 2 , 0 , 0 ... , 0)$$

In this case $T(\Pi_0) = Q(\Pi_0)$ , $R_{c,1} + R_{c,2} = 2R_{c,1}$ and $\gamma_r = e_1 + e_2$ then we will have from Theorem 6

***Lemma 20.-*** The last possible place for unitarity is $B(\Pi_0) = i - 1$ .

From Theorem 7 $A(\Pi_0) = B(\Pi_0)$ then

***Theorem 21.-*** $M_{\Pi_0 + z\varepsilon}$ is unitarizable if and only if $z \leq i - 1$

***Case II***       $\Pi_2 = \Pi_3 = ... = \Pi_{n-1}$    $2\Pi_n = 1$       $\left(\bullet \Pi_0 , \mu_n \circledR = 1\right)$

The root sistem $Q(\Pi_0)$ is

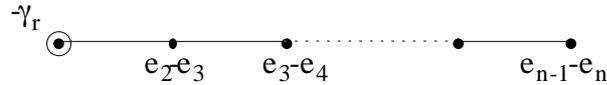

it is of type su $(1, n - 1)$
and the sistem $T(\Pi_0)$

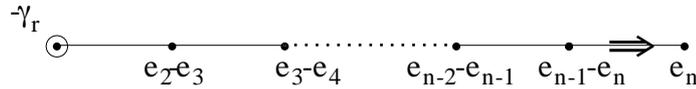

which is of type so $(2n - 1 , 2)$

This case corresponds to the paragraph **a)** of Theorem 6, then

***Lemma 22.*** - The last possible place for unitarity is $B(\Pi_0) = n - 1/2$

From Theorem 7

***Theorem 23.*** - $M_{\Pi_0 + z\varepsilon}$ is unitarizable if and only if $z \leq n - 1/2$

**so (2n - 2 , 2)**

Let $\Lambda = \Pi_0 + z\varepsilon$ , $\Pi_0 = (\Pi_1 , ... , \Pi_n)$ and $\varepsilon = (1, 0, ... , 0)$ ; $\Pi_0$ is $k_1$-dominant and integral, $\Pi_2 \geq \Pi_3 \geq ... \geq \Pi_{n-1} \geq {}^{™}\Pi_n{}^{™}$ and $\Pi_i - \Pi_j \in \mathcal{N}$ for $2 \leq i < j \leq n$ . From the condition $\bullet \Pi_0 + R$ , $\gamma_r \circledR = 0$ it follows $\Pi_1 + \Pi_2 = -2n + 3$ .

We consider two cases

***Case I***  Let the following conditions on the $\Pi_0$ components:

$$\Pi_2 = \Pi_3 = ... = \Pi_i \neq \Pi_{i+1} \qquad i \neq n - 1 \quad \text{o} \quad i = n - 1 \quad \text{y} \quad \Pi_{n-1} \neq -\Pi_n$$

The root sistem $Q(\Pi_0) = T(\Pi_0)$ is

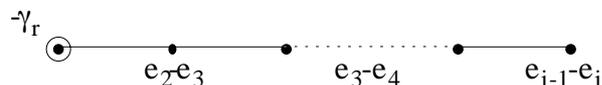

which is of type su $(1 , i - 1)$ . Then by Theorem 6

*Lemma 24.-* The last possible place for unitarity is $B(\Pi_0) = i - 1$

From Theorem 7 $A(\Pi_0) = B(\Pi_0)$, then

*Theorem 25.-* $M_{\Pi_0 + z\varepsilon}$ is unitarizable if and only if $z \le i - 1$

**Case II**  $\qquad \Pi_2 = \Pi_3 = \ldots = \Pi_{n-2} = \Pi_{n-1} = -\Pi_n \qquad (\Pi_n < 0 )$

The root sistem $Q(\Pi_0) = T(\Pi_0)$ is

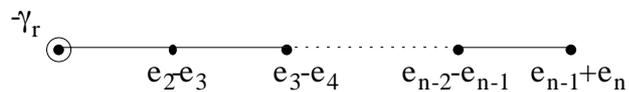

which is of type su $(1, n - 1)$, then

*Theorem 26.-* $M_{\Pi_0 + z\varepsilon}$ is unitarizable if and only if $z \le n - 1$

**e₆**

Let be $\Lambda = \Pi_0 + z\,\varepsilon$ , $\Pi_0 = (\Pi_1, \dots, \Pi_8)$ and $\varepsilon = (0, 0, 0, 0, 0, -2/3, -2/3, 2/3)$ ; $\Pi_0$ is $k_1$-dominant and integral therefore $^{™}\Pi_1{}^{™} \leq \Pi_2 \leq \dots \leq \Pi_5$ , $\Pi_i - \Pi_j \in \mathbb{Z}$ , $2\Pi_i \in \mathbb{Z}$ , $i$ , $j \leq 5$. From the condition $•\Pi_0 + R$ , $\gamma_r \circledR = 0$ we obtain $\Pi_1 + \Pi_2 + \Pi_3 + \Pi_4 + \Pi_5 - \Pi_6 - \Pi_7 + \Pi_8 = -22$

We consider two cases

***Case I*** In the following paragraphs we show, under the conditions given for $\Pi_0$ , that the

root sistems $Q(\Pi_0) = T(\Pi_0)$

   1)  $•\Pi_0$ , $\mu_2 \circledR > 0$

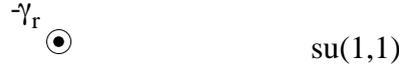
$\mathrm{su}(1,1)$

   2)  $•\Pi_0$ , $\mu_2 \circledR = 0$ , $•\Pi_0$ , $\mu_4 \circledR > 0$

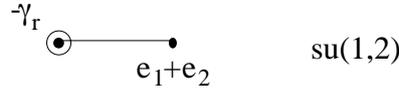
$\mathrm{su}(1,2)$

   3)  $•\Pi_0$ , $\mu_2 \circledR = •\Pi_0$ , $\mu_4 \circledR = 0$ , $•\Pi_0$ , $\mu_5 \circledR > 0$ , $•\Pi_0$ , $\mu_3 \circledR > 0$

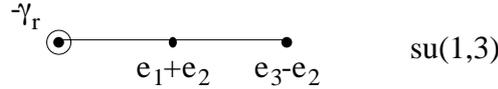
$\mathrm{su}(1,3)$

   4)  (a)  $•\Pi_0$ , $\mu_2 \circledR = •\Pi_0$ , $\mu_4 \circledR = •\Pi_0$ , $\mu_3 \circledR = 0$ , $•\Pi_0$ , $\mu_5 \circledR > 0$

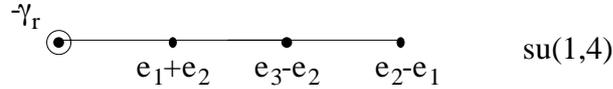
$\mathrm{su}(1,4)$

      (b)  $•\Pi_0$ , $\mu_2 \circledR = •\Pi_0$ , $\mu_4 \circledR = •\Pi_0$ , $\mu_5 \circledR = 0$ , $•\Pi_0$ , $\mu_6 \circledR > 0$    $•\Pi_0$ , $\mu_3 \circledR > 0$

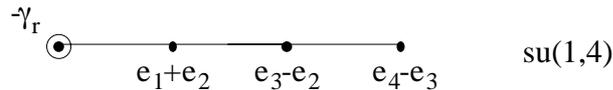
$\mathrm{su}(1,4)$

   5)  $•\Pi_0$ , $\mu_2 \circledR = •\Pi_0$ , $\mu_4 \circledR = •\Pi_0$ , $\mu_5 \circledR = •\Pi_0$ , $\mu_6 \circledR = 0$    $•\Pi_0$ , $\mu_3 \circledR > 0$

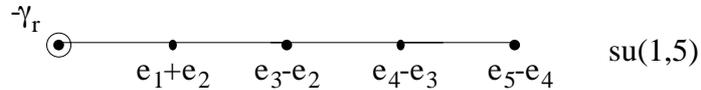
$\mathrm{su}(1,5)$

Then the root sistem $Q(\Pi_0) = T(\Pi_0)$ is of type $\mathrm{su}(1, i)$ with $1 \leq i \leq 5$ and by Theorem 6

***Lemma 27.-*** The last possible place for unitarity is $B(\Pi_0) = i$

By Theorem 7 $A(\Pi_0) = B(\Pi_0)$. Therefore

***Theorem 28.-*** $M_{\Pi_0 + z\varepsilon}$ is unitarizable if and only if $z \leq i$.

***Case II*** Let be the following conditions on the $\Pi_0$ components:

    $•\Pi_0$ , $\mu_2 \circledR = •\Pi_0$ , $\mu_3 \circledR = •\Pi_0$ , $\mu_4 \circledR = •\Pi_0$ , $\mu_5 \circledR = 0$ , $•\Pi_0$ , $\mu_6 \circledR > 0$

The root sistem $Q(\Pi_0) = T(\Pi_0)$ is

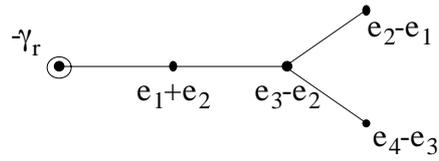

which is of type so $(8, 2)$ . Then $R_{c,1} + R_{c,2} = 2R_{c,1} = 2\ (0, 1, 2, 3, 0, 0, 0, 0)$. In this algebra $\gamma_r = 1/2\ (e_1 + e_2 + e_3 + e_4 + e_5 - e_6 - e_7 + e_8$ ). Therefore we will have by Theorem 6

*Lemma 29.-* The last possible place for unitarity is $B(\Pi_0) = 7$.

In addition $A(\Pi_0) = 7 - (2 - 1) \times 3$

*Lemma 30.-* The first possible place for non unitarity is $A(\Pi_0) = 4$

*Theorem 31.-* $M_{\Pi_0 + z\varepsilon}$ is unitarizable if and only if $z \le 4$ or $z = 7$

**e₇**

Let $\Lambda = \Pi_0 + z\varepsilon$ , $\Pi_0 = (\Pi_1, \ldots, \Pi_8)$ with $\Pi_7 = -\Pi_8$ and
$\varepsilon = (0, 0, 0, 0, 0, 1, -1/2, 1/2)$ ; $\Pi_0$ is $k_1$ -dominant and integral therefore
$^{™}\Pi_1{}^{™} \le \Pi_2 \le \ldots \le \Pi_5$ , $\Pi_i - \Pi_j \in Z$ , $2\Pi_i \in Z$ , $1 \le i \le j \le 5$ , and
$1/2\ (\Pi_1 - \Pi_2 - \Pi_3 - \Pi_4 - \Pi_5 - \Pi_6 - \Pi_7 + \Pi_8) \in \mathcal{N}$

The condition $\bullet \Pi_0 + R$ , $\gamma_r \circledR = 0$ fix the value $\Pi_8 = -17/2$ . In all the cases

$$\mu_1 = 1/2\ (e_1 - e_2 - e_3 - e_4 - e_5 - e_6 - e_7 + e_8).$$

We consider two cases

*Case I*

Given the conditions on $\Pi_0$ we can calculate the root sistems $Q(\Pi_0) = T(\Pi_0)$.

1) $\bullet \Pi_0 , \mu_1 \circledR > 0$

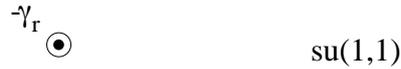

su(1,1)

2) $\bullet \Pi_0 , \mu_1 \circledR = 0$ , $\bullet \Pi_0 , \mu_3 \circledR > 0$

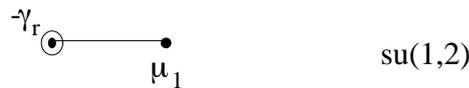

su(1,2)

3) $\bullet \Pi_0 , \mu_1 \circledR = \bullet \Pi_0 , \mu_3 \circledR = 0$ , $\bullet \Pi_0 , \mu_4 \circledR > 0$

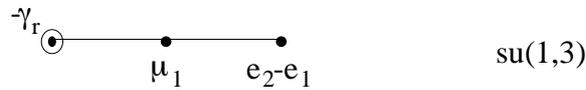

su(1,3)

4) $\bullet \Pi_0 , \mu_1 \circledR = \bullet \Pi_0 , \mu_3 \circledR = \bullet \Pi_0 , \mu_4 \circledR = 0$ , $\bullet \Pi_0 , \mu_2 \circledR > 0$ , $\bullet \Pi_0 , \mu_5 \circledR > 0$

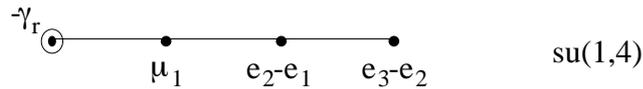

su(1,4)

5) (a) $\bullet \Pi_0 , \mu_1 \circledR = \bullet \Pi_0 , \mu_3 \circledR = \bullet \Pi_0 , \mu_4 \circledR = \bullet \Pi_0 , \mu_2 \circledR = 0$ , $\bullet \Pi_0 , \mu_5 \circledR > 0$

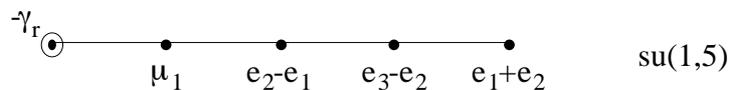

su(1,5)

(b) $\bullet \Pi_0 , \mu_1 \circledR = \bullet \Pi_0 , \mu_3 \circledR = \bullet \Pi_0 , \mu_4 \circledR = \bullet \Pi_0 , \mu_5 \circledR = 0$ , $\bullet \Pi_0 , \mu_2 \circledR > 0$ , $\bullet \Pi_0 , \mu_6$ $\circledR > 0$

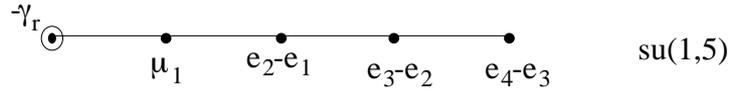

su(1,5)

6) $\bullet \Pi_0 , \mu_1 \circledR = \bullet \Pi_0 , \mu_3 \circledR = \bullet \Pi_0 , \mu_4 \circledR = \bullet \Pi_0 , \mu_5 \circledR = \bullet \Pi_0 , \mu_6 \circledR = 0$ , $\bullet \Pi_0 , \mu_2 \circledR$ > 0

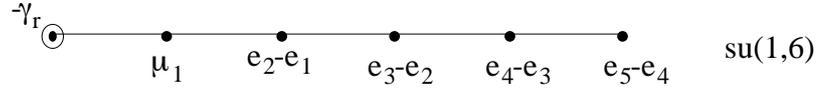

su(1,6)

Then the root sistem $Q(\Pi_0) = T(\Pi_0)$ is of type su $(1 , i)$ with $1 \le i \le 6$ and by Theorem 6

*Lemma 32.-* The last possible place for unitarity is $B(\Pi_0) = i$

By Theorem 7 $A(\Pi_0) = B(\Pi_0)$

*Theorem 33.-* $M_{\Pi_0 + z\varepsilon}$ is unitarizable if and only if $z \le i$

*Case II* Given the following conditions on the $\Pi_0$ components:

$\bullet \Pi_0 , \mu_1 \circledR = \bullet \Pi_0 , \mu_2 \circledR = \bullet \Pi_0 , \mu_3 \circledR = \bullet \Pi_0 , \mu_4 \circledR = \bullet \Pi_0 , \mu_5 \circledR = 0$ , $\bullet \Pi_0 , \mu_6 \circledR >$ 0

The root sistem $Q(\Pi_0) = T(\Pi_0)$ is

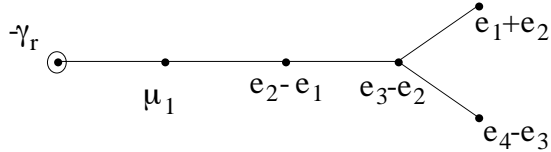

and it is of type so $(10 , 2)$. Then

$R_{c,1} + R_{c,2} = 2R_{c,1} = 2 (0, 1, 2, 3, -2, -2, -2, 2)$. For this algebra $\gamma_r = e_8 - e_7$ then we will have from Theorem 6

*Lemma 34.-* The last possible place for unitarity is $B(\Pi_0) = 9$.

On the other hand $A(\Pi_0) = 9 - (2 -1) \times 4 = 5$ then

*Theorem 35.-* $M_{\Pi_0 + z\varepsilon}$ is unitarizable if and only if $z \le 5$ or $z = 9$.

## VII.- POSSIBLE PLACES FOR UNITARITY

We consider two cases:

### A) Cases with more than one reduction point

**su(p,q)**

Dynkin diagram 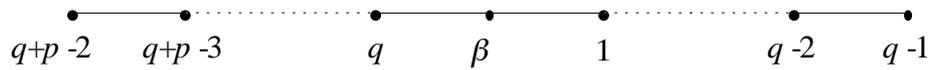

Let $M_\Lambda$ be a representation for su(p, q) with $\Lambda = (\Lambda_1 , \Lambda_2 , ... , \Lambda_{p+q})$ with respect to the standard orthonormal basis of $R^n$ $(n = p + q)$ , satisfying the following conditions on its components:

$$\Lambda_1 = \Lambda_2 = ... = \Lambda_i \ne \Lambda_{i+1}$$

$$\Lambda_n = \Lambda_{n-1} = ... = \Lambda_{n-j+1} \ne \Lambda_{n-j}$$

If we put $\Lambda = \Lambda_0 + \varepsilon\lambda$ these conditions are equivalent to the following ones:

$\bullet \Lambda_0 , \mu_{q-1} \circledR = \bullet \Lambda_0 , \mu_{q-2} \circledR = .... = \bullet \Lambda_0 , \mu_{t+1} \circledR = 0$ , $\bullet \Lambda_0 , \mu_t \circledR \ne 0$

$\bullet \Lambda_0 , \mu_{n-2} \circledR = \bullet \Lambda_0 , \mu_{n-3} \circledR = ... = \bullet \Lambda_0 , \mu_{s+1} \circledR = 0$ , $\bullet \Lambda_0 , \mu_s \circledR \ne 0$

with $t = q - j$ and $s = n - i - 1$

Applying steps **i**) and **ii**) of Jakobsen method (section IV) we obtain (see Fig. 5).

$$\alpha_0 = \beta + \mu_1 + ... + \mu_t + \mu_q + \mu_{q+1} + ... + \mu_s$$

such that $n_{\alpha_0} = t + s - q + 2$ . Then $\lambda_0 = q - t - s - 1$

Then a first order polynomial will be missing with highest weight

$$\Lambda_0 + ( q - t - s - 1 )\varepsilon + R - \alpha_0$$

where, in this case, $\varepsilon = (q/n , q/n , ... , q/n , -p/n , ... , -p/n )$ with $p$ copies of $q/n$ and $q$ of $-p/n$ .

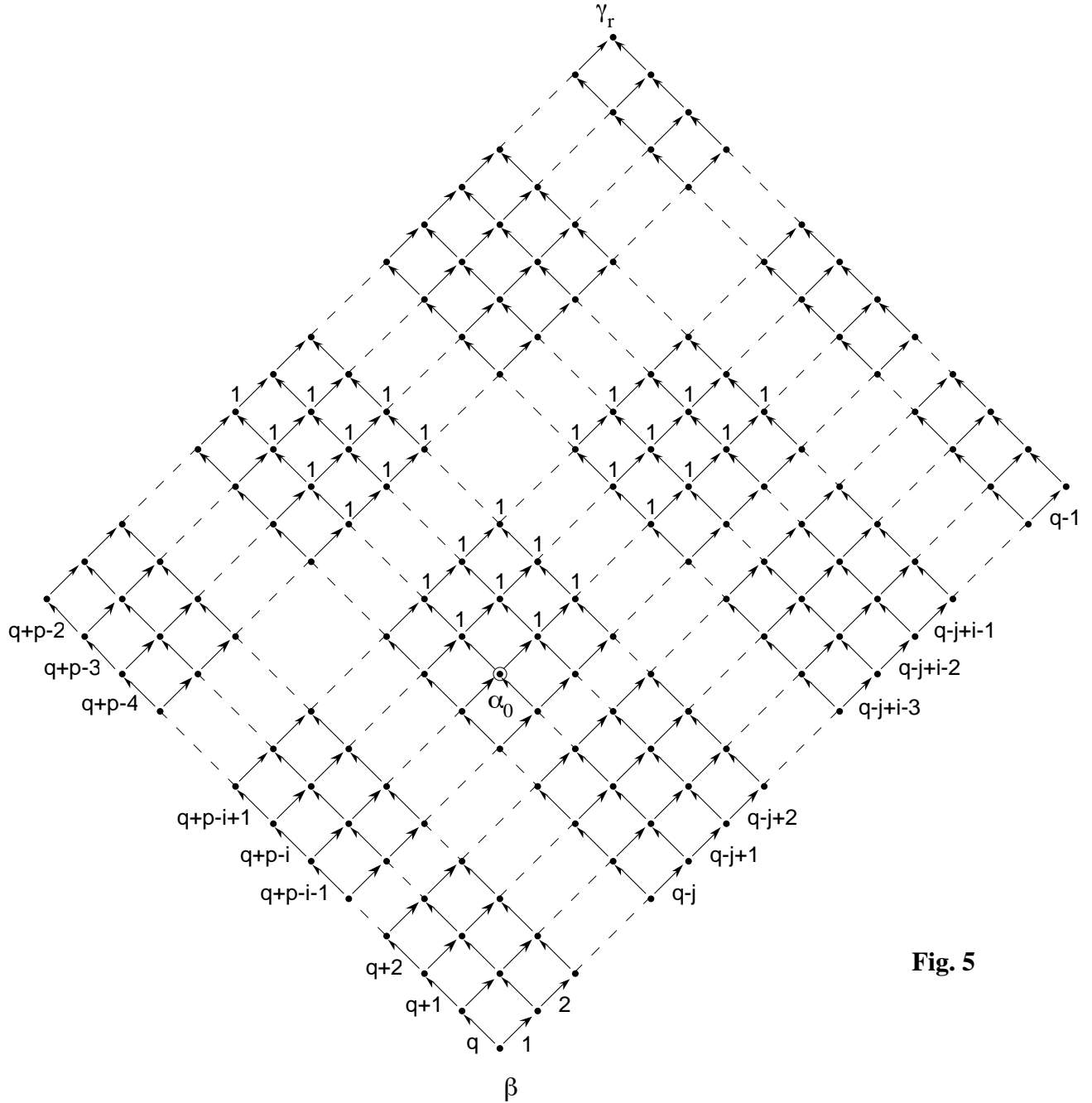

**Fig. 5**

For $\lambda_q = \lambda_0 + \lambda_s = \lambda_0 - 1$ we obtain from steps **iii**) and **iv**) a second order polynomial that will be missing with heighest weight

$$\Lambda_0 + (\lambda_0 - 2)\varepsilon + R - \alpha^1_{2-\lambda_0} - \alpha^2_{2-\lambda_0}$$

Next case is $\lambda_q = \lambda_0 - 2$ where a third order polynomial with highest weight

$\Lambda_0 + (\lambda_0 - 2)\varepsilon + R - \alpha_{3-\alpha_0} - \alpha_{3-\lambda_0} - \alpha_{3-\lambda_t}$ will be missing. Continuing in the same way we arrive at $\lambda_q = \lambda_0 - u$ , $u = \min (q - t - 1, n - s - 2)$ where a polynomial of order $u + 1 = \min (i, j)$ will be missing. For $\lambda < \lambda_0 - u$ it is impossible to find a polynomial of order strictly higher than $\min (i, j)$ because the $k_1$-dominance is violated (see step **iv**).

Thus the reduction level is $\min (i, j)$. On the other hand $\lambda = \bullet \Lambda$, $\gamma_r \circledR$ or,

equivalently, $\Lambda_n = \Lambda_1 - \lambda$ . Then, taking into account the possible values of $\lambda$ we obtain the following diagram:

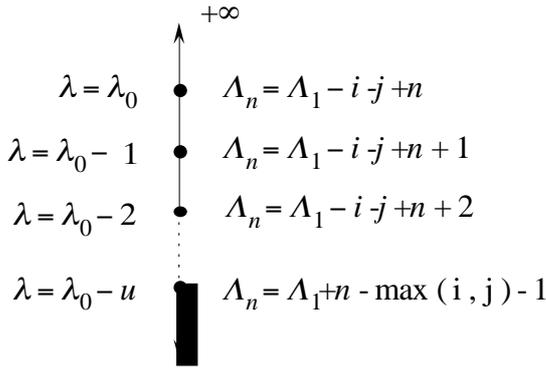

Now we apply the EHW method:

$\Lambda = \Pi_0 + z\,\varepsilon$  with  $\Pi_0 = (\Pi_1, \dots, \Pi_n)$

From  $\langle \Pi_0 + R, \gamma_r \rangle = 0$  we obtain

$$\Pi_1 - \Pi_n + n - 1 = 0$$

The root sistem $Q(\Pi_0)$ is of type $su(i, j)$ then the last possible place for unitarity, following Lemma 8  corresponds to  $z = i + j - 1$  and we will have :

$$\Lambda_1 = \Pi_1 + (i + j - 1)q/n \quad , \quad \Lambda_n = \Pi_1 + n - 1 - (i + j - 1)p/n$$

or what is the same :  $\Lambda_n = \Lambda_1 - i - j + n$ .

The next place for unitarity corresponds to  $z = i + j - 2$  for which  $\Lambda_n = \Lambda_1 - i - j + n + 1$. Continuing along this way and having in mind that the first reduction point corresponds to (Lemma 9) :  $z = \max(i, j)$  we arrive at the first possible place for non unitarity for which:

$$\Lambda_1 = \Pi_1 + \max(i, j)\, q/n \quad ; \quad \Lambda_n = \Pi_1 + n - 1 - \max(i, j) p/n$$

then  $\Lambda_n = \Lambda_1 - \max(i, j) - 1 + n$ . We obtain, as desired, the same results as in Jakobsen method.

## sp (n, $\mathcal{R}$)

Dynkin diagram 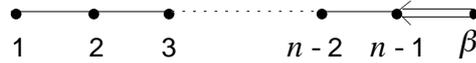

Let $M_\Lambda$ be a representation for sp (n, $\mathcal{R}$) with $\Lambda = (\Lambda_1, \dots, \Lambda_n)$ we put $\Lambda = \Lambda_0 + \lambda\varepsilon$ where $\varepsilon = (1, 1, \dots, 1)$. We consider two cases:

### Case I

The weight $\Lambda$ satisfy the following conditions on its components

$$\Lambda_1 = \Lambda_2 = \dots = \Lambda_i \geq \Lambda_{i+1} + 2$$

or, equivalently

$$\langle \Lambda_0, \mu_1 \rangle = \langle \Lambda_0, \mu_2 \rangle = \dots = \langle \Lambda_0, \mu_{i-1} \rangle = 0 \ , \ \langle \Lambda_0, \mu_i \rangle = n \geq 2$$

Applying Jakobsen method we obtain (see Fig. 6)

$$\alpha_0 = \beta + 2\mu_{n-1} + 2\mu_{n-2} + \dots + 2\mu_{n-(n-i)}$$

with $\Pi_{\alpha_0} = 2(n - i) + 1$ . As $\alpha_0$ is a long root, the condition  $\langle \Lambda + R, \alpha_0 \rangle = 1$ implies

$$\frac{1}{2}(\Lambda_0, \alpha_0) + \lambda_0 + n - i + 1 = 1 \quad , \quad \lambda_0 = i - n$$

then a first order polynomial with highest weight  $\Lambda_0 + (i - n)\varepsilon + R - \alpha_0$  will be missing when we unitarize.

For $\lambda = \lambda_0 + \lambda_s = i - n - \frac{2}{2}$  we obtain  a second order polynomial which will be missing with highest weight

$$\Lambda_0 + (i - n - \tfrac{1}{2})\varepsilon + R - 2\,\alpha^1{}_{2(n-i)+2}$$

For $\lambda = \lambda_0 + 2\lambda_s$ the third order missing polynomial has highest weight

$$\Lambda_0 + (i - n - 1)\varepsilon + R - 2\,\alpha^1{}_{2(n-i)+3} - \alpha^2{}_{2(n-i)+3}$$

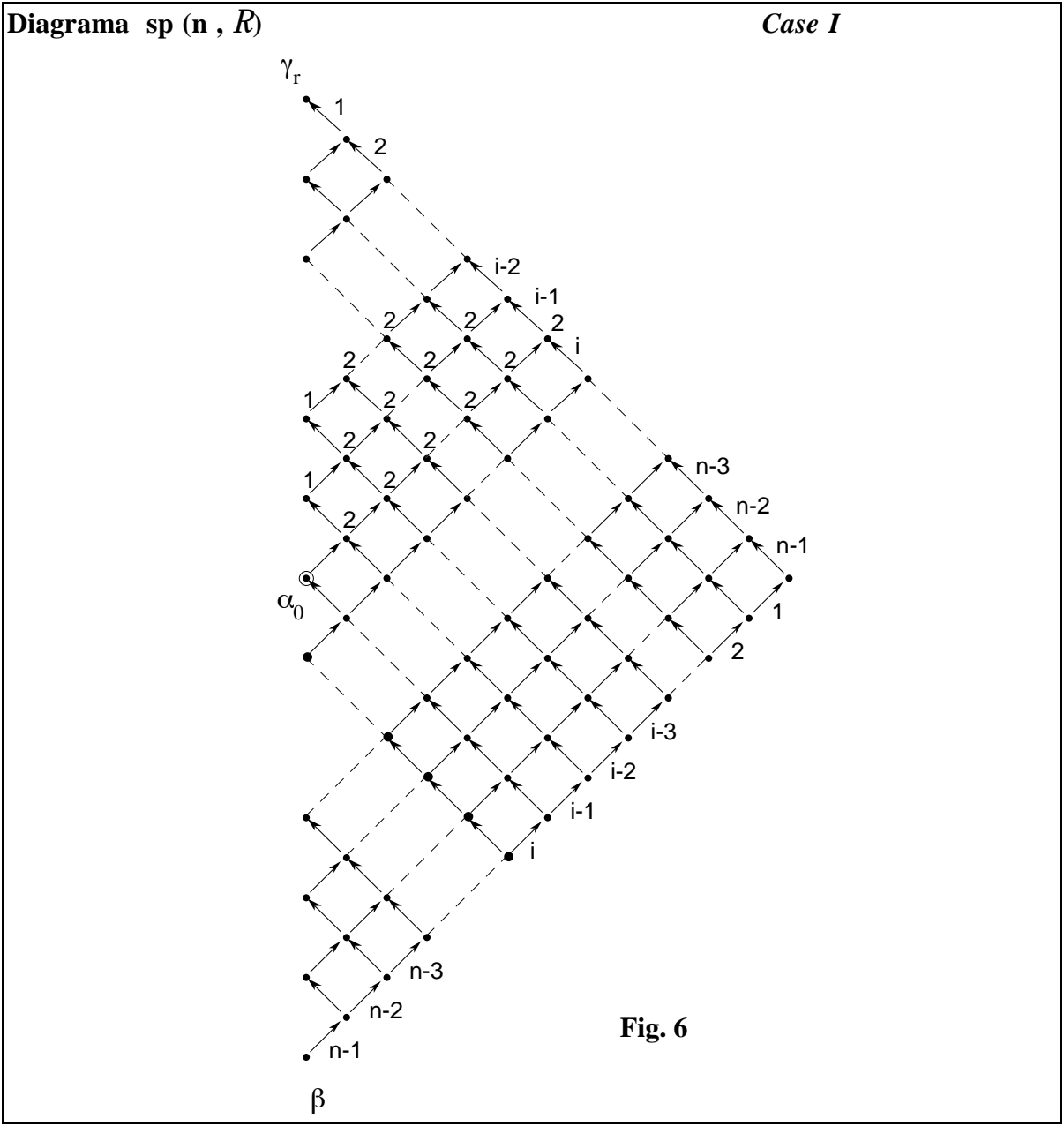

**Diagrama  sp (n , $\overline{R}$)**                                    *Case I*

**Fig. 6**

Following along these lines we arrive at  $\lambda = \lambda_0 + (i-1)\lambda_s = \tfrac{1}{2}(i+1) - n$  where a i-th order polynomial will be missing. For  $\lambda < \tfrac{1}{2}(i+1) - n$  it is impossible to obtain polynomials of order strictly higher than $i$  because there would not be $k_1$-dominance, therefore the reduction level is $i$ . On the other hand from the condition  $\lambda = \bullet \Lambda, \gamma_r \circledR$  it follows that  $\Lambda_1 = \lambda$  Thus, for the different values of $\lambda$  we obtain the following diagram which give us the possible values of $\Lambda_1$ for unitarity:

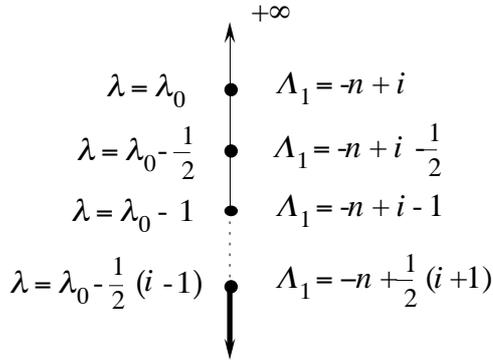

We now apply the EHW method
$$\Lambda = \Pi_0 + z\varepsilon \quad , \quad \Pi_0 = (\Pi_1, \ldots, \Pi_n) \quad , \quad \Pi_1 = -n$$

The root sistem $\ Q(\Pi_0)\ $ is of type $\mathrm{sp}\,(i,$
$\mathcal{R})$ then the last possible place for unitarity
corresponds to $z = i$ following Lemma 11,
then
$$\Lambda_1 = \Pi_1 + i \quad ; \quad \Lambda_1 = -n + i$$

The next place corresponds to $\ z = i - \frac{}{2}\ $,

therefore $\ \Lambda_1 = -n + i - \frac{}{2}\ $. Continuing along this way we obtain for the first reduction point (see Lemma 12) $z = \frac{}{2}(i+1)\ $ hence the first component of the weight is $\Lambda_1 = -n + \frac{}{2}(i+1)$, as desired.

## Case II

We consider in this case the following conditions
$$\Lambda_1 = \Lambda_2 = \ldots = \Lambda_i \quad ; \quad \Lambda_i - \Lambda_{i+1} = 1 \quad , \quad \Lambda_{i+1} = \Lambda_{i+2} = \ldots = \Lambda_{i+j} \neq \Lambda_{i+j+1}$$

which are equivalent to the following ones:

$$\bullet \Lambda_0, \mu_1 \circledR = \bullet \Lambda_0, \mu_2 \circledR = \ldots = \bullet \Lambda_0, \mu_{i-1} \circledR = 0 \quad , \quad \bullet \Lambda_0, \mu_i \circledR = 1$$

$$\bullet \Lambda_0, \mu_{i+1} \circledR = \bullet \Lambda_0, \mu_{i+2} \circledR = \ldots = \bullet \Lambda_0, \mu_{i+j-1} \circledR = 0 \quad , \quad \bullet \Lambda_0, \mu_{i+j} \circledR = n \ \geq 1$$

In this case (see Fig. 7 and 8) applying Jakobsen method
$$\alpha_0 = \beta + 2\left(\mu_{n-1} + \mu_{n-2} + \ldots + \mu_{n-(n-i-j)}\right) + \mu_{n-(n-i-j)} + \ldots + \mu_{n-(n-i)}$$

| **Diagrama sp (n , $\mathcal{R}$)** | ***Caso II*** |
|---|---|
| | **a)** $j = i+1$ |

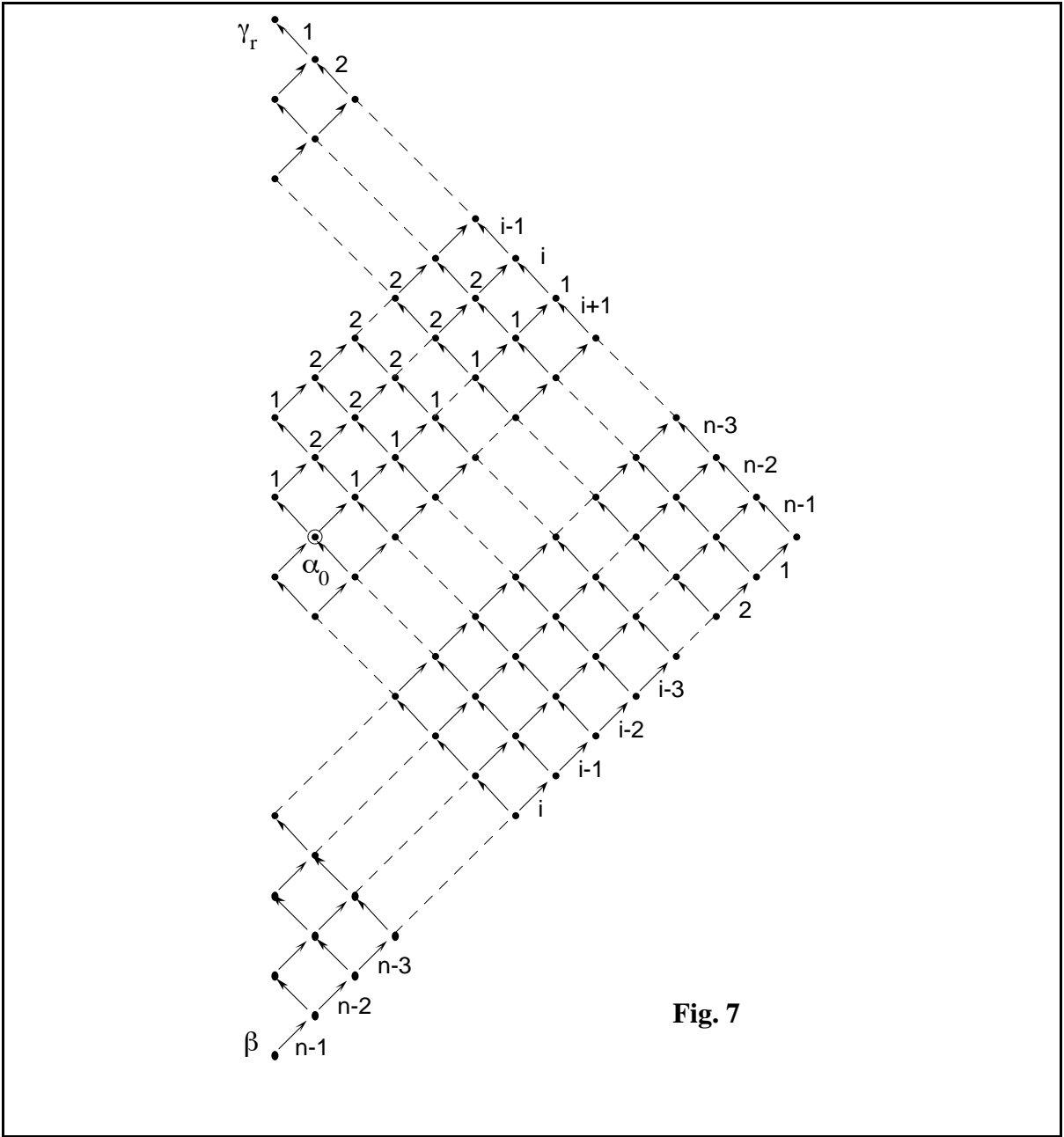

**Fig. 7**

| Diagrama sp (n , $\bar{R}$) | *Caso II* | b) $j > i+1$ |
|---|---|---|

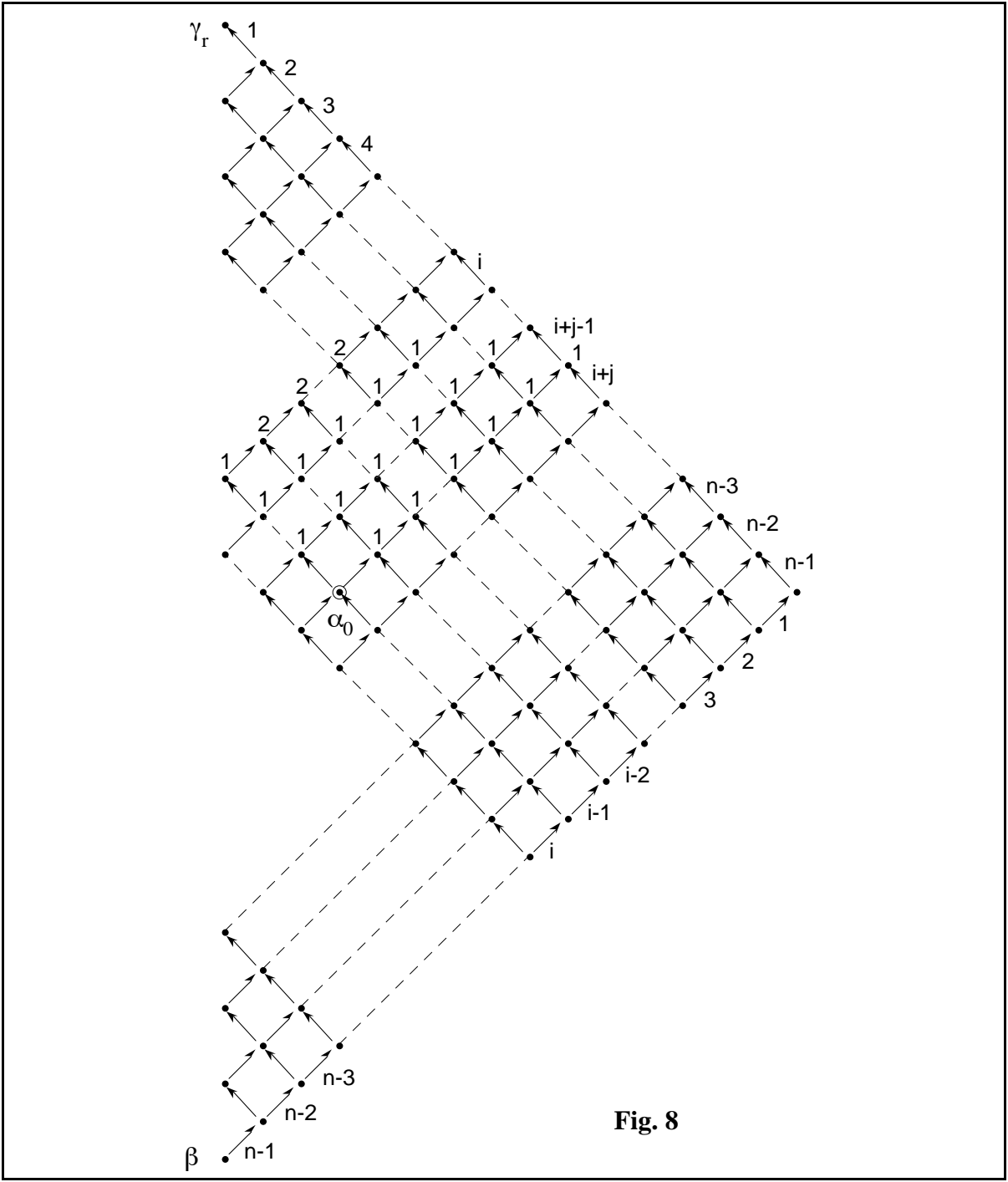

**Fig. 8**

the height of which is $H_{\alpha_0} = 2(n-i) - j + 1$. As $\alpha_0$ is a short root, the condition $\bullet \Lambda + R$, $\alpha_0 \circledR = 1$ implies

$$(\Lambda_0, \alpha_0) + 2\lambda_0 + 2(n-i+1) - j = 1$$

and, having in mind that we can also state

$$\alpha_0 = \gamma_r - 2(\mu_1 + \ldots + \mu_{i-1}) - \mu_i - \mu_{i+1} - \ldots - \mu_{i+j-1}$$

we have

$$\lambda_0 = i - n + \frac{j}{2}$$

For $\lambda = \lambda_0 + \lambda_s = i - n + \frac{\mu_{-}}{2}$ we obtain a second order polynomial which will be missing with highest weight

$$\Lambda_0 + \left( i - n + \frac{(j-1)}{2} \right)\varepsilon + R - \alpha^1_{2(n-i+1)-j} - \alpha^2_{2(n-i+1)-j}$$

Continuing as in case I we arrive at $\lambda = \lambda_0 + (i-1)\lambda_s = -n + \frac{1}{2}(i+j+1)$ where a i-th order polynomial will be missing. Since $\lambda < -n + \frac{1}{2}(i+j+1)$ there is no $k_1$-dominance therefore the reduction level is $i$ . The diagram in this case is the following:

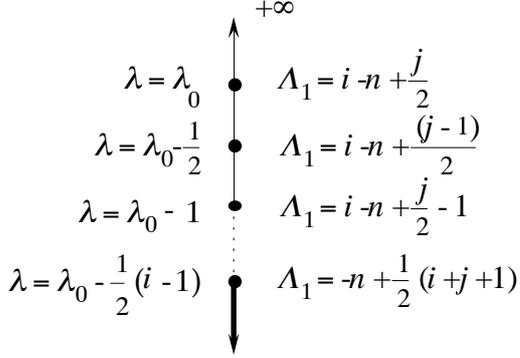

We now apply the EHW method

The root system $Q(\Pi_0)$ is of type sp $(i, \leftarrow)$ and $T(\Pi_0)$ is of type sp $(k, \leftarrow)$ with $k = i + j$ Then the last possible place for unitarity corresponds, by Lemma 14, to $z = \frac{-}{2}(2i+j)$ therefore in this case $\Lambda_1 = -n + i + \frac{z}{2}$.

Following along this way we obtain the first possible place for non unitarity $\Lambda_1 = -n + \frac{-}{2}(i+j+1)$ by Lemma 15.

**so\* (2n)**

Dynkin diagram 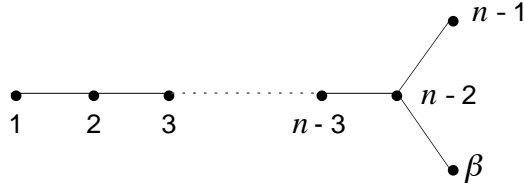

Let now $M_\Lambda$ be a representation for so\*(2n) with $\Lambda = (\Lambda_1, ... , \Lambda_n)$. In this case, given $\Lambda = \Lambda_0 + \lambda\varepsilon$ we have $\varepsilon = (\frac{-}{2}, \frac{-}{2}, ... , \frac{-}{2})$. We consider the following conditions on its components

$$\Lambda_1 = \Lambda_2 = ... = \Lambda_i > \Lambda_{i+1} + 1 \qquad i \neq 1$$

or, equivalently,

$$\bullet \Lambda_0, \mu_1 \circledR = \bullet \Lambda_0, \mu_2 \circledR = .... = \bullet \Lambda_0, \mu_{i-1} \circledR = 0 \qquad , \qquad \bullet \Lambda_0, \mu_i \circledR = n > 1$$

From the Jakobsen method we have (see Fig. 9)

$$\alpha_0 = \beta + (\mu_{n-1} + \mu_{n-2}) + (\mu_{n-2} + \mu_{n-3}) + ... + (\mu_{n-(n-i)} + \mu_{n-(n-i)-1})$$

with $\pi_{\alpha_0} = 2(n-i) + 1$. The condition $\bullet \Lambda + R , \alpha_0 \circledR = 1$ implies in this case

$$(\Lambda_0, \alpha_0) + \lambda_0 + 2(n-i) + 1 = 1 \qquad ; \qquad \lambda_0 = 2(i-n)$$

For $\lambda = \lambda_0 + \lambda_s = 2(i-n-1)$ we obtain a second order polynomial which will be missing with highest weight

$$\Lambda_0 + 2(i-n-1)\varepsilon + R - \alpha^1_{2(n-i)+3} - \alpha^2_{2(n-i)+3}$$

Following along these lines we arrive at $\lambda = \lambda_0 + \left\{ [\frac{l}{2}] - 1 \right\}\lambda_s$ where a polynomial with order $[\frac{l}{2}]$ is missing. For $\lambda < 2\left( i - n - \left\{ [\frac{l}{2}] - 1 \right\} \right)$ there is impossible to obtain missing polynomials the order of which is strictly higher than $[\frac{l}{2}]$ because in those places the weights are not $k_1$-dominants therefore the reduction level is $[\frac{l}{2}]$. From the condition $\lambda = \bullet \Lambda , \gamma_r \circledR$ we obtain $\Lambda_1 = \frac{z}{2}$. In this way we obtain the following diagram

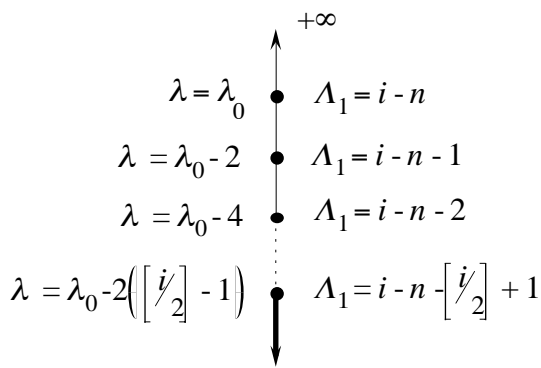

Now we apply EHW method

$$\Lambda = \Pi_0 + z\varepsilon \quad , \quad \Pi_0 = (\Pi_1, \ldots, \Pi_n) \quad ,$$

$$2\Pi_1 = -2n + 3$$

The root system is $Q(\Pi_0) = T(\Pi_0)$ which is of type so*($2i$), therefore, following Lemma 17, the last possible place for unitarity is $z = 2i - 3$, and at this place

$$\Lambda_1 = \Pi_1 + \frac{1}{2}(2i - 3) = i - n$$

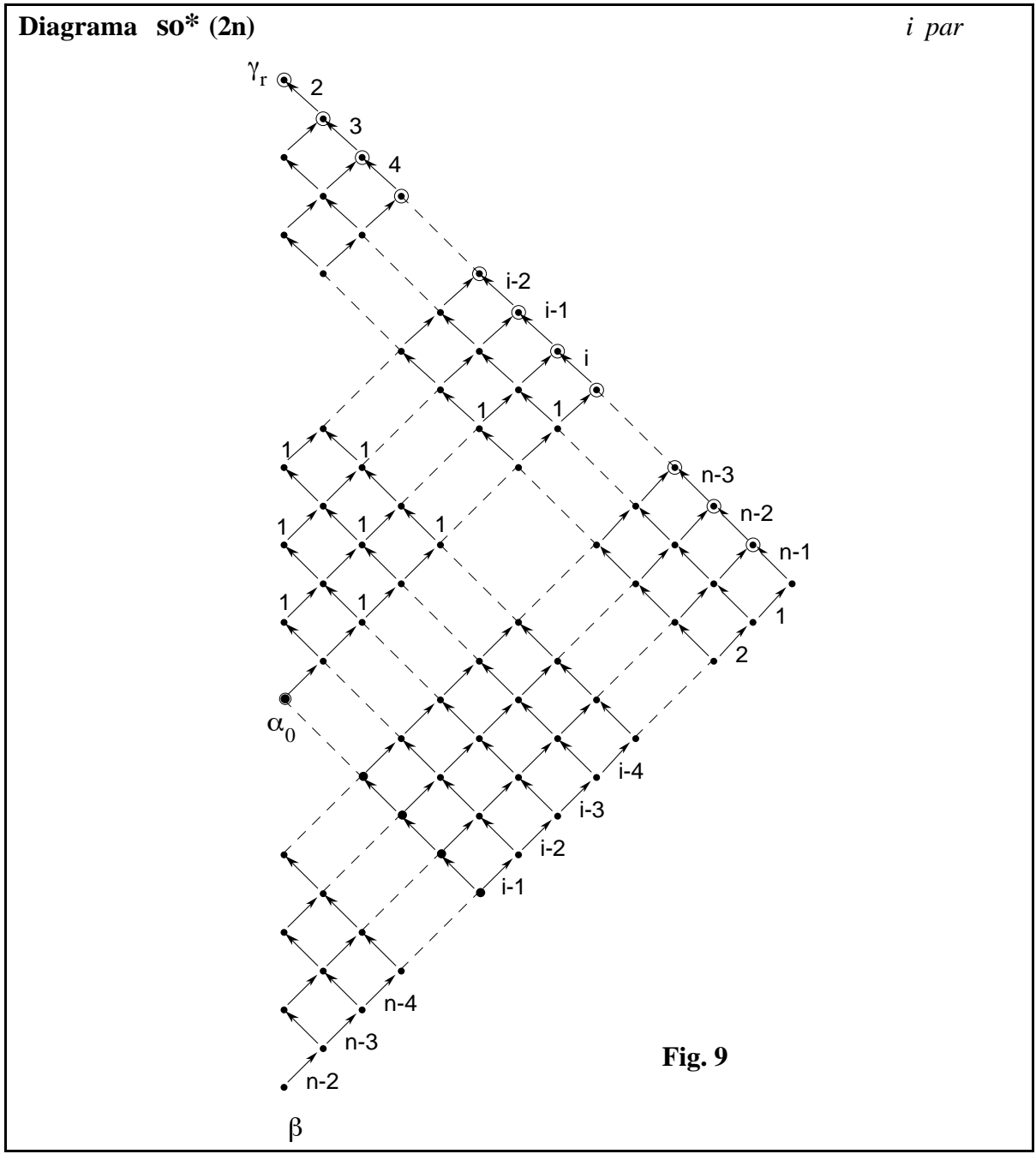

**Diagrama so\* (2n)**                                    *i par*

**Fig. 9**

The next place corresponds to $z = 2i - 5$ then in this case $\Lambda_1 = i - n - 1$. Continuing along this way we arrive at the first reduction point where by Lemma 18 $z =$ $\begin{bmatrix} i-1 \text{ if } i \text{ is even} \\ i \text{ if } i \text{ is odd} \end{bmatrix}$ or, what is the same

$$z = 2i - 1 - 2\left[\tfrac{i}{2}\right]$$

At this place $\Lambda_1 = \Pi_1 + \frac{1}{2}\left(2i - 1 - 2\left[\tfrac{i}{2}\right]\right) = i - n - \left(\left[\tfrac{i}{2}\right] - 1\right)$, as desired.

Let now be $i = 1$ and the following conditions:

$$\Lambda_1 \neq \Lambda_2 = \Lambda_3 = \ldots = \Lambda_j = \Lambda_{j+1} \neq \Lambda_{j+2}$$

or what is the same

$\bullet\Lambda_0, \mu_1 \circledast = m_1 > 0$    $\bullet\Lambda_0, \mu_2 \circledast = \ldots = \bullet\Lambda_0, \mu_j \circledast = 0$    $\bullet\Lambda_0, \mu_{j+1} \circledast = m_{j+1} > 0$

with $2 \leq j + 1 \leq n - 1$. In this case we have $H_{\alpha_0} = 2n - j - 2$ (in Fig. 9 the possible $\alpha_0$ are inside a little circle on the right branch of $C^-_{\gamma_r}$ ).

From condition $\bullet\Lambda + R$ , , $\alpha_0\circledast = 1$ we obtain $\lambda_0 = 3 - 2n + j$. We will have only one reduction point for each $j$ because $C^+_{\alpha_0}$ is unidimensional:

$$\lambda = \lambda_0 \qquad \Lambda_1 = \frac{3 - 2n + j}{2}$$

If we consider the remarks following Theorem 19 we may see that the above results are the same that those obtained by the EHW method.

### $e_6$

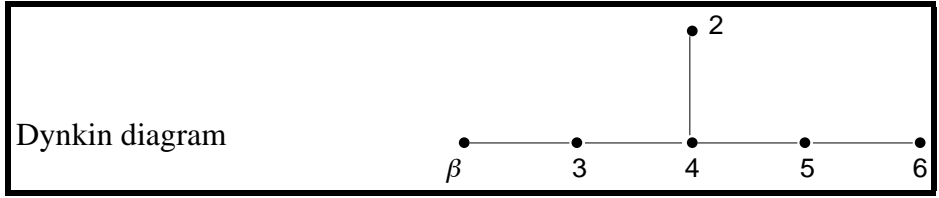

Dynkin diagram

Here $\Lambda = \Lambda_0 + \lambda\varepsilon$ with $\varepsilon = \left(0, 0, 0, 0, 0, -\tfrac{1}{3}, -\tfrac{1}{3}, -\tfrac{1}{3}\right)$ For $\Lambda_0 \neq 0$ the only case for which the reduction level is strictly higher than one (all the rest are excluded for $k_1$-dominance arguments ) is the following

$$\bullet\Lambda_0, \mu_i \circledast = 0 \qquad 2 \leq i \leq 5$$

$$\bullet\Lambda_0, \mu_6\circledast = n > 0$$

Applying Jakobsen method we obtain

$$\alpha_0 = \beta + \mu_3 + \mu_4 + \mu_5 + \mu_6$$

with height 5 (see Fig. 10). From the condition $\bullet\Lambda + R, \alpha_0 \circledast = 1$ we obtain

$1 = \bullet\Lambda, \alpha_0\circledast + \bullet R, \alpha_0\circledast = \bullet\Lambda, \gamma_r \circledast + 5 = \lambda_0 + 5$

; $\lambda_0 = -4$

For $\lambda_q = \lambda_0 + \lambda_s = -7$

$$\left\langle \Lambda + R, \ \alpha_8^1 \right\rangle = \left\langle \Lambda + R, \ \alpha_8^2 \right\rangle = 1$$

then a second order polynomial will be missing with highest weight:

$$\Lambda_0 - 7\varepsilon + R - \alpha_8^1 - \alpha_8^2$$

The value $\lambda_q = -7$ is the first possible place for non unitarity.

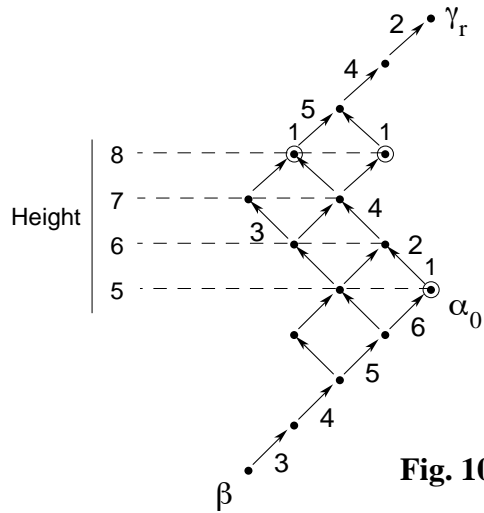

**Fig. 10**

The diagram is then

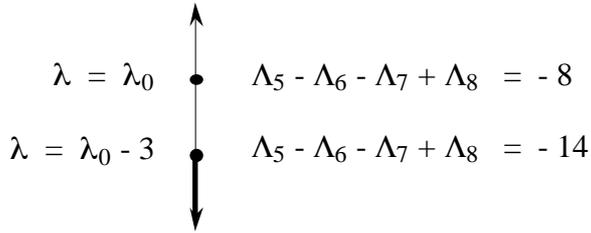

$\lambda = \lambda_0$    $\Lambda_5 - \Lambda_6 - \Lambda_7 + \Lambda_8 = -8$

$\lambda = \lambda_0 - 3$    $\Lambda_5 - \Lambda_6 - \Lambda_7 + \Lambda_8 = -14$

We now apply EHW method:

$\Lambda = \Pi_0 + z\varepsilon$ , $\Pi_0 = (\Pi_1, \dots, \Pi_8)$ ,

$$\Pi_7 = -\Pi_8 = \frac{17}{2}$$

We observe that in the last possible place for unitarity $\Lambda = \Pi_0 + 7\varepsilon$ then

$\Lambda_5 - \Lambda_6 - \Lambda_7 + \Lambda_8 = -22 + 2 \times 7 = -8$ (we see, from the conditions on $\Lambda_0$ that $\Lambda_1 = \Lambda_2 = \Lambda_3 = \Lambda_4 = 0$) and in the first for non unitarity

$$\Lambda_5 - \Lambda_6 - \Lambda_7 + \Lambda_8 = -22 + 2 \times 4 = -14$$

**$e_7$**

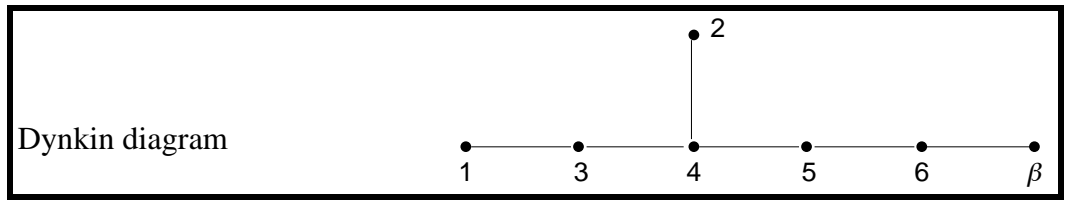

Dynkin diagram

For this case $\varepsilon = \left(0, 0, 0, 0, 0, 1, -\dfrac{1}{2}, \dfrac{1}{2}\right)$. As in $e_6$ we must only consider one case:

$$\bullet \Lambda_0, \mu_i \circledR = 0 \qquad 1 \le i \le 5$$

$$\bullet \Lambda_0, \mu_6 \circledR = n > 0$$

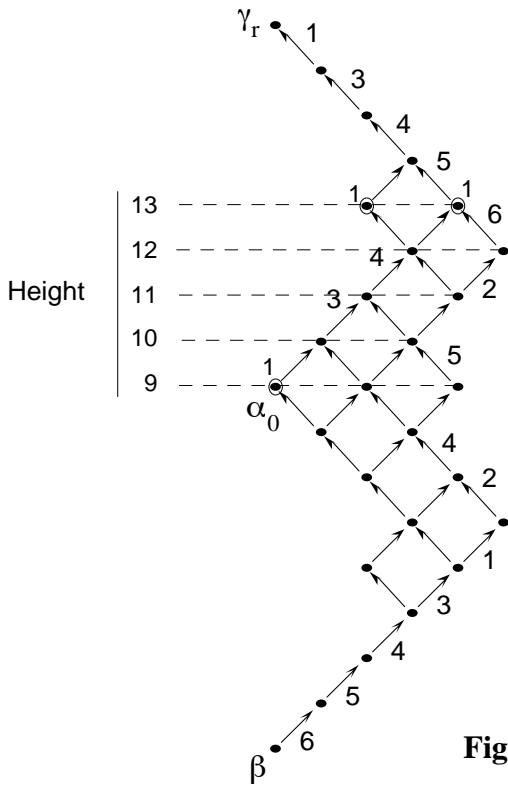

Fig. 11

with those conditions (see Fig. 11) the Jakobsen method gives

$$\alpha_0 = \beta + 2\mu_6 + 2\mu_5 + 2\mu_4 + \mu_3 + \mu_2$$

with height 9, then

$1 = \bullet \Lambda, \alpha_0 \circledR + \bullet R, \alpha_0 \circledR = \bullet \Lambda, \alpha_0 \circledR + 9 = \lambda_0 + 9$ ;   $\lambda_0 = -8$

For $\lambda_q = -12$

$$\left\langle \Lambda + R, \alpha_{13}^1 \right\rangle = \left\langle \Lambda + R, \alpha_{13}^2 \right\rangle = 1$$

and we obtain, for $\lambda_q = -12$ , a missing second order polynomial with highest weight

$$\Lambda_0 - 12\varepsilon + R - \alpha_{13}^1 - \alpha_{13}^2$$

The weight diagram is, then

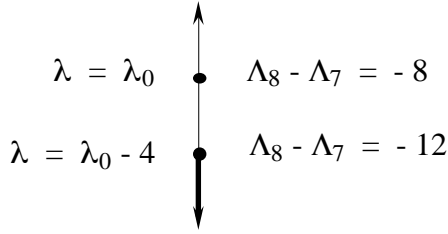

$\lambda = \lambda_0$      $\Lambda_8 - \Lambda_7 = -8$

$\lambda = \lambda_0 - 4$      $\Lambda_8 - \Lambda_7 = -12$

We now apply EHW method $\Lambda = \Pi_0 + z\varepsilon$ , $\Pi_0 = (\Pi_1, \Pi_2, \ldots, \Pi_8)$ with $\Pi_1 + \Pi_2 + \Pi_3 + \Pi_4 + \Pi_5 - \Pi_6 - \Pi_7 + \Pi_8 = -22$. We observe that in the last possible place for unitarity (Lemma 34)

$\Lambda = \Pi_0 + 9\varepsilon$   $\Lambda_8 = \Pi_8 + \dfrac{-}{2}$   $\Lambda_7 = \Pi_7 - \dfrac{-}{2}$

then $\Lambda_8 - \Lambda_7 = -8$ and in the first for non unitarity by Theorem 35 , $\Lambda = \Pi_0 + 5\varepsilon$ then $\Lambda_8 - \Lambda_7 = -12$ as desired.

## B) Cases with one single reduction point

In the following we want to consider the cases with one single reduction point that are not included in the general scheme studied in case **A**). The missing highest weights are all of the form $\Lambda - \alpha_0$.

**e₆**

With the conditions given for $\Lambda_0$, the values for $\alpha_0$ (see Fig. 12) and $\lambda_0$ are the following ones ( we observe that the equation $\langle \Lambda + R \, , \, \alpha_0 \rangle = 1$ implies $\Lambda_0 = 1 - H_{\alpha_0}$ ):

1)    •$\Lambda_0$ , $\mu_2 \circledR > 0$

       $\alpha_0 = \gamma_r$

       $\lambda_0 = -10$

2)    •$\Lambda_0$ , $\mu_2 \circledR = 0$   •$\Lambda_0$ , $\mu_4 \circledR > 0$

       $\alpha_0 = \gamma_r - \mu_2$

       $\lambda_0 = -9$

3)    •$\Lambda_0$ , $\mu_2 \circledR = $ •$\Lambda_0$ , $\mu_4 \circledR = 0$

       •$\Lambda_0$ , $\mu_5 \circledR > 0$   •$\Lambda_0$ , $\mu_3 \circledR > 0$

       $\alpha_0 = \gamma_r - \mu_2 - \mu_4$

       $\lambda_0 = -8$

4) a) •$\Lambda_0$ , $\mu_2 \circledR = $ •$\Lambda_0$ , $\mu_4 \circledR = $ •$\Lambda_0$ , $\mu_3 \circledR = 0$

       •$\Lambda_0$ , $\mu_5 \circledR > 0$

       $\alpha_0 = \gamma_r - \mu_2 - \mu_4 - \mu_3$

       $\lambda_0 = -7$

   (b)   •$\Lambda_0$ , $\mu_2 \circledR = $ •$\Lambda_0$ , $\mu_4 \circledR = $ •$\Lambda_0$ , $\mu_5 \circledR = 0$ ,    •$\Lambda_0$ , $\mu_6 \circledR > 0$    ,    •$\Lambda_0$ , $\mu_3 \circledR > 0$

       $\alpha_0 = \gamma_r - \mu_2 - \mu_4 - \mu_5$

       $\lambda_0 = -7$

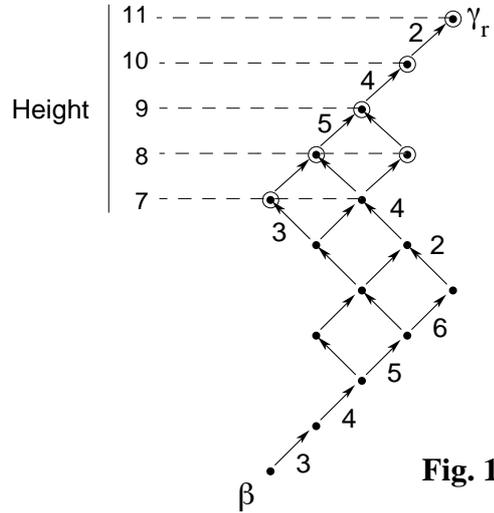

**Fig. 12**

5)    $\bullet \Lambda_0 , \mu_2 \circledR = \bullet \Lambda_0 , \mu_4 \circledR = \bullet \Lambda_0 , \mu_5 \circledR = \bullet \Lambda_0 , \mu_6 \circledR = 0$        $\bullet \Lambda_0 , \mu_3 \circledR > 0$

$\alpha_0 = \gamma_r - \mu_2 - \mu_4 - \mu_5 - \mu_6$

$\lambda_0 = -6$

Given $\Lambda = ( \Lambda_1 , \Lambda_2 , \dots , \Lambda_8 )$ from the condition $\langle \Lambda , \gamma_r \rangle = \lambda$ we obtain

$$\Lambda_1 + \Lambda_2 + \Lambda_3 + \Lambda_4 + \Lambda_5 - \Lambda_6 - \Lambda_7 + \Lambda_8 = 2\lambda$$

Then we will have the following diagram for the conditions i) , $1 \leq i \leq 5$

$$\lambda = \lambda_0 \quad\updownarrow\quad \Lambda_1 + \Lambda_2 + \Lambda_3 + \Lambda_4 + \Lambda_5 - \Lambda_6 - \Lambda_7 + \Lambda_8 \;=\; -22 + 2i$$

Now let $\Lambda = \Pi_0 + z\varepsilon$ , as in EHW method for the algebra $e_6$ in A).

We have, following Theorem 28 that for the paragraph i) , $1 \leq i \leq 5$

$$\Lambda_1 + \Lambda_2 + \Lambda_3 + \Lambda_4 + \Lambda_5 - \Lambda_6 - \Lambda_7 + \Lambda_8 = -22 + 2i$$

## $e_7$

For this algebra (Fig. 13) holds also

$\lambda_0 = 1 - H_{\alpha_0} :$

1)    $\bullet \Lambda_0 , \mu_1 \circledR > 0$

$\alpha_0 = \gamma_r$

$\lambda_0 = -16$

2)    $\bullet \Lambda_0 , \mu_1 \circledR = 0$      $\bullet \Lambda_0 , \mu_3 \circledR > 0$

$\alpha_0 = \gamma_r - \mu_1$

$\lambda_0 = -15$

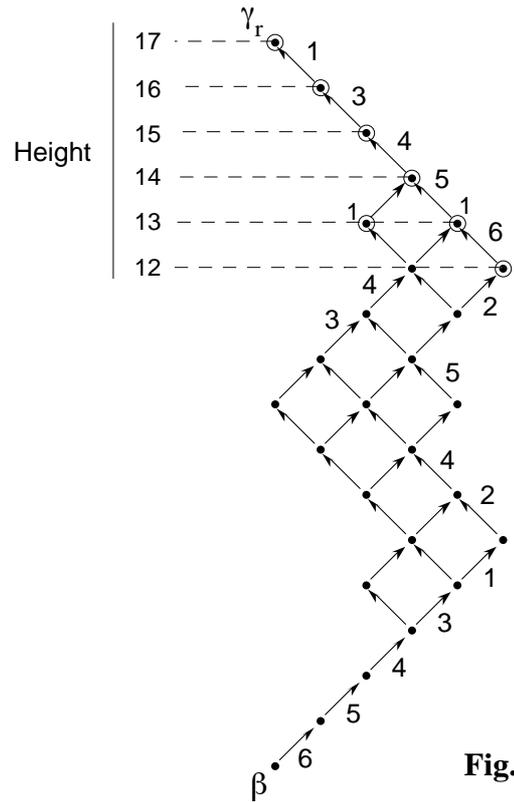

Fig. 13

3)    $\bullet \Lambda_0 , \mu_1 \circledR = \bullet \Lambda_0 , \mu_3 \circledR = 0$ ,   $\bullet \Lambda_0 , \mu_4 \circledR$

$> 0$

$\alpha_0 = \gamma_r - \mu_1 - \mu_3$

$\lambda_0 = -14$

4)    $\bullet \Lambda_0 , \mu_1 \circledR = \bullet \Lambda_0 , \mu_3 \circledR = \bullet \Lambda_0 , \mu_4 \circledR = 0$

$\bullet \Lambda_0 , \mu_2 \circledR > 0$      $\bullet \Lambda_0 , \mu_5 \circledR > 0$

$\alpha_0 = \gamma_r - \mu_1 - \mu_3 - \mu_4$

$\lambda_0 = -13$

5) (a) $\bullet \Lambda_0 , \mu_1 \circledR = \bullet \Lambda_0 , \mu_3 \circledR = \bullet \Lambda_0 , \mu_4 \circledR =$

$= \bullet \Lambda_0 , \mu_2 \circledR = 0$    ,    $\bullet \Lambda_0 , \mu_5 \circledR > 0$

$\alpha_0 = \gamma_r - \mu_1 - \mu_3 - \mu_4 - \mu_2$

$\lambda_0 = -12$

(b)  $\bullet \Lambda_0 , \mu_1 \circledR = \bullet \Lambda_0 , \mu_3 \circledR = \bullet \Lambda_0 , \mu_4 \circledR = \bullet \Lambda_0 , \mu_5 \circledR = 0$ ,   $\bullet \Lambda_0 , \mu_2 \circledR > 0$ ,  $\bullet \Lambda_0 , \mu_6 \circledR$

$> 0$

$$\alpha_0 = \gamma_r - \mu_1 - \mu_3 - \mu_4 - \mu_5$$
$$\lambda_0 = -12$$

6)  $\langle {}^\bullet\Lambda_0, \mu_1 \circledR = \langle {}^\bullet\Lambda_0, \mu_3 \circledR = \langle {}^\bullet\Lambda_0, \mu_4 \circledR = \langle {}^\bullet\Lambda_0, \mu_5 \circledR = \langle {}^\bullet\Lambda_0, \mu_6 \circledR = 0$ ,  $\langle {}^\bullet\Lambda_0, \mu_2 \circledR$

$> 0$

$$\alpha_0 = \gamma_r - \mu_1 - \mu_3 - \mu_4 - \mu_5 - \mu_6$$
$$\lambda_0 = -11$$

Here also  $\Lambda = ( \Lambda_1 , \Lambda_2 , \dots , \Lambda_8 )$  and from  $\langle \Lambda , \gamma_r \rangle = \lambda$  we have  $\Lambda_8 - \Lambda_7 = \lambda$ .

Thus the diagram for the paragraph i) ,  $1 \le i \le 6$

$$\lambda = \lambda_0 \quad \updownarrow \quad \Lambda_8 - \Lambda_7 = -17 + i$$

is

Given  $\Lambda = \Pi_0 + z \varepsilon$  as in the EHW method for this algebra in A) we have following Theorem 33

$$\Lambda_8 = \Pi_8 + \frac{i}{2} \quad , \quad \Lambda_7 = \Pi_7 - \frac{i}{2}$$

therefore  $\Lambda_8 - \Lambda_7 = -17 + i$  for the paragraph i),  $1 \le i \le 6$.

## so (2n - 1, 2)

We consider two cases

### *Case I*

Given the following conditions

$\langle {}^\bullet\Lambda_0, \mu_2 \circledR = \langle {}^\bullet\Lambda_0, \mu_3 \circledR = \dots = \langle {}^\bullet\Lambda_0, \mu_{i-1} \circledR = 0$   $\langle {}^\bullet\Lambda_0, \mu_i \circledR > 0$ ,  $i < n$  or  $i = n$  and  $\langle {}^\bullet\Lambda_0, \mu_n \circledR > 1$  which are equivalent to the following ones

$$\Lambda_2 = \Lambda_3 = \dots = \Lambda_{i-1} = \Lambda_i > \Lambda_{i+1} \qquad \Lambda_n = \frac{1}{2} \quad \text{if} \quad i = n$$

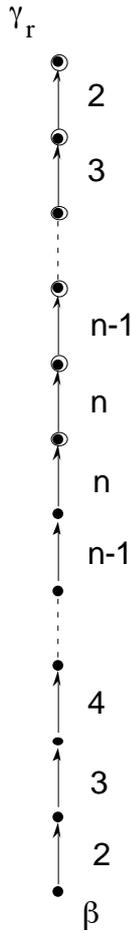

**Fig. 14**

In this case $\alpha_0 = \gamma_r - \mu_2 - \mu_3 - \ldots - \mu_{i-1}$ (Fig. 14) and $\lambda_0 = i - 2n + 1$.

Given
$$\Lambda = (\Lambda_1, \Lambda_2, \ldots, \Lambda_n)$$
from
$$\langle \Lambda, \gamma_r \rangle = \Lambda_1 + \Lambda_2 = \lambda$$
we obtain the diagram:

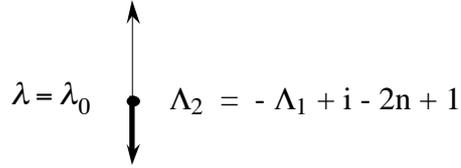

$$\lambda = \lambda_0 \qquad \Lambda_2 = -\Lambda_1 + i - 2n + 1$$

Using EHW method we obtain, following Theorem 21
$$\left(\Lambda_1, \Lambda_2, \ldots, \Lambda_n\right) = \left(\Pi_1, \Pi_2, \ldots, \Pi_n\right) + (i-1)(1, 0, \ldots, 0)$$
and having in mind that $\Pi_1 + \Pi_2 = -2n + 2$ we have
$$\Lambda_1 + \Lambda_2 = i - 2n + 1$$

### Case II

Let now be the following conditions on $\Lambda_0$ :

$$\bullet \Lambda_0, \mu_2 \circledR = \bullet \Lambda_0, \mu_3 \circledR = \ldots = \bullet \Lambda_0, \mu_{n-1} \circledR = 0 \quad, \quad \bullet \Lambda_0, \mu_n \circledR = 1$$

which are equivalent to the following ones:
$$\Lambda_2 = \Lambda_3 = \ldots = \Lambda_n = \frac{1}{2}$$
In this case $\alpha_0 = \gamma_r - \mu_2 - \ldots - \mu_n$ (Fig. 14) and $\lambda_0 = -n + \frac{3}{2}$

The weight diagram is now

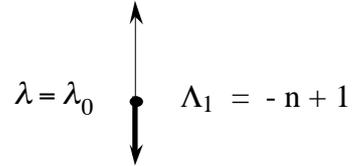

$$\lambda = \lambda_0 \qquad \Lambda_1 = -n + 1$$

If we apply EHW method we obtain
$$\left(\Lambda_1, \Lambda_2, \ldots, \Lambda_n\right) = \left(\Pi_1, \Pi_2, \ldots, \Pi_n\right) + \left(n - \frac{1}{2}\right)(1, 0, \ldots, 0)$$

therefore $\Lambda_1 + \Lambda_2 = -n + \frac{3}{2}$ and having in mind that $\Lambda_2 = \Lambda_n = \frac{1}{2}$ then $\Lambda_1 = -n + 1$ .

**so (2n - 2, 2)**

### Case I

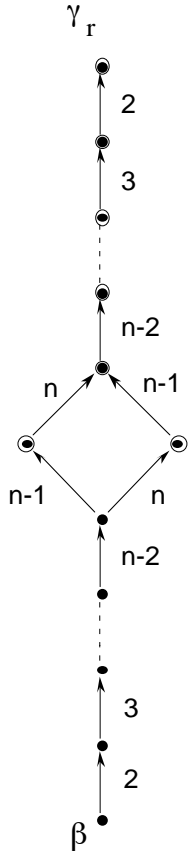

$\gamma_r$

n-2

n    n-1

n-1    n

n-2

$\beta$

**Fig. 15**

Assume that

$\bullet \Lambda_0 , \mu_2 \circledR = \bullet \Lambda_0 , \mu_3 \circledR = ... = \bullet \Lambda_0 , \mu_{i-1} \circledR = 0$

$\bullet \Lambda_0 , \mu_i \circledR > 0$ with $i \neq n - 1$ or $i = n - 1$ and

$\bullet \Lambda_0 , \mu_n \circledR > 0$

which are equivalent to the following ones ( here also $\Lambda = ( \Lambda_1 , \Lambda_2 , ... , \Lambda_n )$ )

$\Lambda_2 = \Lambda_3 = \cdots = \Lambda_{i-1} = \Lambda_i > \Lambda_{i+1}$   $i \neq n - 1$ or

$\Lambda_2 = \Lambda_3 = \cdots = \Lambda_n \geq \frac{1}{2}$   respectively.

With this conditions we obtain from Jakobsen method

$$\alpha_0 = \gamma_r - \mu_2 - \mu_3 - \cdots - \mu_{i-1}$$

(Fig. 15) and $\lambda_0 = -2n + i + 2$ .

The weight diagram is    $\lambda = \lambda_0$  $\updownarrow$   $\Lambda_1 + \Lambda_2 = i - 2n + 2$

Applying the EHW method we obtain, following Theorem 25

$$\left( \Lambda_1 , \Lambda_2 , ... , \Lambda_n \right) = \left( \Pi_1 , \Pi_2 , ... , \Pi_n \right) + (i - 1)(1 , 0 , ... , 0)$$

and having in mind that $\Pi_1 + \Pi_2 = -2n + 3$   we have

$$\Lambda_1 + \Lambda_2 = i - 2n + 2 .$$

*Case II*

$\bullet \Lambda_0 , \mu_2 \circledR = \bullet \Lambda_0 , \mu_3 \circledR = ... = \bullet \Lambda_0 , \mu_{n-2} \circledR = \bullet \Lambda_0 , \mu_n \circledR = 0$    ,    $\bullet \Lambda_0 , \mu_{n-1} \circledR > 0$

which are equivalent to the following ones:

$$\Lambda_2 = \Lambda_3 = \dots = \Lambda_{n-1} > \Lambda_n   \text{and}   \Lambda_{n-1} = -\Lambda_n$$

In this case we have

$\alpha_0 = \gamma_r - \mu_2 - \mu_3 - \cdots - \mu_{n-2} - \mu_n$   and   $\lambda_0 = 2 - n$

The weight diagram is then    $\lambda = \lambda_0$  $\updownarrow$  $\Lambda_1 + \Lambda_2 = 2 - n$

Applying the EHW method we obtain

$$\left( \Lambda_1 , \Lambda_2 , ... , \Lambda_n \right) = \left( \Pi_1 , \Pi_2 , ... , \Pi_n \right) + (n - 1)(1 , 0 , ... , 0)$$

by Theorem 26. Therefore   $\Lambda_1 + \Lambda_2 = 2 - n$

We complete in this way the equivalence between Jakobsen method and EHW method.

## VIII. UNITARY REPRESENTATIONS OF SU(2,2)

The algebra su (2 , 2) has Dynkin diagram

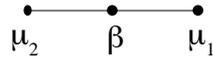

and Jakobsen diagram

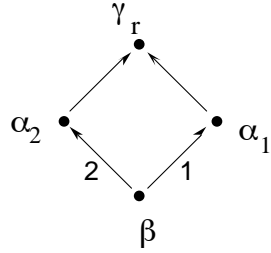

with

$\mu_1 = e_3 - e_4$ , $\mu_2 = e_1 - e_2$ , $\beta = e_2 - e_3$ , $\gamma_r = e_1 - e_4$ , $\alpha_1 = e_2 - e_4$ , $\alpha_2 = e_1 - e_3$

Given a weight $\Lambda$ we may decompose it in the following way

$$\Lambda = \Lambda_0 + \lambda \varepsilon \qquad \text{with} \qquad \varepsilon = \frac{1}{2} \ (1, \ 1, \ -1, \ -1)$$

We consider the following cases:

## Case I

Let be the following conditions on the $\Lambda_0$ components:

        i) $\bullet \Lambda_0$ , $\mu_1 \circledR = 0$

        ii) $\bullet \Lambda_0$ , $\mu_2 \circledR = n > 0$

Having in mind that $\bullet \Lambda_0$ , $\gamma_r \circledR = 0$ we obtain

$$\Lambda_0 = \left( \frac{n}{4}, \frac{-3n}{4}, \frac{n}{4}, \frac{n}{4} \right)$$

For this case the last possible place for unitarity is equal to the first one for non unitarity and it corresponds to $\lambda_0 = -1$ :

$$\Lambda = \left( \frac{n-2}{4}, \frac{-3n-2}{4}, \frac{n+2}{4}, \frac{n+2}{4} \right)$$

We have $\alpha_0 = e_1 - e_3$ (Fig 16) then the weight that we must exclude (missing weight) in order to obtain unitarity is

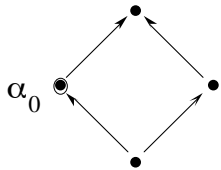

$$\Lambda - \alpha_0 = \left( \frac{n-6}{4}, \frac{-3n-2}{4}, \frac{n+6}{4}, \frac{n+2}{4} \right)$$

**Fig. 16**

## Case II

        i) $\bullet \Lambda_0$ , $\mu_1 \circledR = n > 0$

        ii) $\bullet \Lambda_0$ , $\mu_2 \circledR = 0$

In this case

$$\Lambda_0 = \left( \frac{-n}{4}, \frac{-n}{4}, \frac{3n}{4}, \frac{-n}{4} \right)$$

Here also $\lambda_0 = -1$ , then

$$\Lambda = \left( \frac{-n-2}{4}, \frac{-n-2}{4}, \frac{3n+2}{4}, \frac{-n+2}{4} \right)$$

Having in mind that $\alpha_0 = e_2 - e_4$ (Fig 17) the missing weight will be

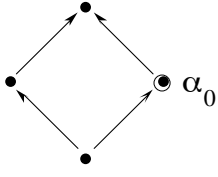

$$\Lambda - \alpha_0 = \left( \frac{-n-2}{4} , \frac{-n-6}{4} , \frac{3n+2}{4} , \frac{-n+6}{4} \right)$$

**Fig. 17**

***Case III.***

i) • $\Lambda_0$ , $\mu_1 \circledR = n > 0$

ii) • $\Lambda_0$ , $\mu_2 \circledR = m > 0$

With this conditions

$$\Lambda_0 = \left( \frac{m-n}{4} , \frac{-3m-n}{4} , \frac{m+3n}{4} , \frac{m-n}{4} \right)$$

in this case $\lambda_0 = -2$ then

$$\Lambda = \left( \frac{m-n-4}{4} , \frac{-3m-n-4}{4} , \frac{m+3n+4}{4} , \frac{m-n+4}{4} \right)$$

As we can see in Fig 18 in this case $\alpha_0 = \gamma_r$. The missing weight will be

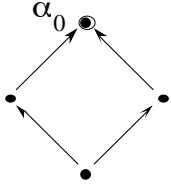

$$\Lambda - \alpha_0 = \left( \frac{m-n-8}{4} , \frac{-3m-n-4}{4} , \frac{m+3n+4}{4} , \frac{m-n+8}{4} \right)$$

**Fig. 18**

Having in mind the above results we can stablish the set of unitary representations (we remind that $n$ , $m \in \mathcal{N}$ and $\lambda \in \leftarrow$) for su $(2, 2)$:

a) $\left( \frac{n-2}{4} , \frac{-3n-2}{4} , \frac{n+2}{4} , \frac{n+2}{4} \right)$   $\forall n > 0$

Extremal vector: $Y = E_{-\beta} E_{-\mu_2} + n E_{-\alpha_2}$   ,   $\Lambda_Y = \Lambda - \alpha_2$

b) $\left( \frac{n+2\lambda}{4} , \frac{-3n+2\lambda}{4} , \frac{n-2\lambda}{4} , \frac{n-2\lambda}{4} \right)$   $\forall n > 0$ , $\forall \lambda < -1$

c) $\left( \frac{-n-2}{4} , \frac{-n-2}{4} , \frac{3n+2}{4} , \frac{-n+2}{4} \right)$   $\forall n > 0$

Extremal vector: $Y = E_{-\beta} E_{-\mu_1} - n E_{-\alpha_1}$   ,   $\Lambda_Y = \Lambda - \alpha_1$

d) $\left( \frac{-n+2\lambda}{4} , \frac{-n+2\lambda}{4} , \frac{3n-2\lambda}{4} , \frac{-n-2\lambda}{4} \right)$   $\forall n > 0$ , $\forall \lambda < -1$

e) $\left( \frac{m-n-4}{4} , \frac{-3m-n-4}{4} , \frac{m+3n+4}{4} , \frac{m-n+4}{4} \right)$   $\forall$ $m$ , $n > 0$

Extremal vector:

$Y = E_{-\beta} E_{-\mu_1} E_{-\mu_2} - n E_{-\alpha_1} E_{-\mu_2} + m E_{-\alpha_2} E_{-\mu_1} - mn E_{-\gamma_r}$   ,   $\Lambda_Y = \Lambda - \gamma_r$

f) $\left( \frac{m-n+2\lambda}{4} , \frac{-3m-n+2\lambda}{4} , \frac{m+3n-2\lambda}{4} , \frac{m-n-2\lambda}{4} \right)$   $\forall$ $m$ , $n > 0$   $\forall \lambda < -2$

The restriction to the maximal compact subgroup $SU(2) \times SU(2) \times U(1)$ is given by the generators (LO86):

$$J^{(1)}_{\pm} = E_{\pm\mu_2} \quad , \quad J^{(1)}_3 = \frac{1}{2}(H_1 - H_2)$$

$$J^{(2)}_{\pm} = E_{\pm\mu_1} \quad , \quad J^{(2)}_3 = \frac{1}{2}(H_3 - H_4)$$

$$R_0 = \frac{1}{2}(H_1 + H_2 - H_3 - H_4)$$

Denoting the eigenvalues of $J^{(1)}_3$, $J^{(2)}_3$ and $R_0$ by $j_1, j_2$ y $d$ respectively we have the following correspondence with the components of the highest weights in the standard orthonormal basis:

$$j_1 = \frac{1}{2}(\Lambda_1 - \Lambda_2) \qquad j_2 = \frac{1}{2}(\Lambda_3 - \Lambda_4) \qquad d = \frac{1}{2}(\Lambda_1 + \Lambda_2 - \Lambda_3 - \Lambda_4)$$

In the following we give the unitary representations of $su(2,2)$ in the form $(j_1, j_2; d)$ showing its Poincaré content (MA77).

a) $\left(\frac{n}{2}, 0; \frac{-n-2}{2}\right)$    $n = 1, 2, 3, \dots$
   Missing weight: $\left(\frac{n-1}{2}, \frac{1}{2}; \frac{-n-4}{2}\right)$
   Mass $= 0$   ,    Helicity $\stackrel{\cdot}{=} \frac{n}{2}$

b) $\left(\frac{n}{2}, 0; \frac{-n+2\lambda}{2}\right)$    $n = 1, 2, 3, \dots$ , $-\infty < \lambda < -1$
   Mass $\neq 0$   ,    Spin $= \frac{n}{2}$

c) $\left(0, \frac{n}{2}; \frac{-n-2}{2}\right)$    $n = 1, 2, 3, \dots$
   Missing weight ; $\left(\frac{1}{2}, \frac{n-1}{2}; \frac{-n-4}{2}\right)$
   Mass $= 0$   ,    Helicity $\stackrel{\cdot}{=} \frac{n}{2}$

d) $\left(0, \frac{n}{2}; \frac{-n+2\lambda}{2}\right)$    $n = 1, 2, 3, \dots$ , $-\infty < \lambda < -1$
   Mass $\neq 0$   ,    Spin $= \frac{n}{2}$

e) $\left(\frac{m}{2}, \frac{n}{2}; \frac{-m-n-4}{2}\right)$    $n, m = 1, 2, 3, \dots$
   Missing weight: $\left(\frac{m-1}{2}, \frac{n-1}{2}; \frac{-m-n-6}{2}\right)$    $n, m = 1, 2, 3, \dots$
   Mass $\neq 0$   ,    Spin $\stackrel{\cdot}{=} \frac{m+n}{2}$

f) $\left(\frac{m}{2}, \frac{n}{2}; \frac{-m-n+2\lambda}{2}\right)$    $n, m = 1, 2, 3, \dots$ , $-\infty < \lambda < -2$
   Mass $\neq 0$   ,    Spin $= \left|\frac{n-m}{2}\right|, \left|\frac{n-m}{2}\right| + 1, \dots, \left|\frac{n+m}{2}\right|$

## IX. UNITARY REPRESENTATIONS OF SO(3,2)

The Dynkin diagram of the de Sitter algebra $so(3,2)$ is 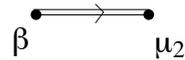

and its Jakobsen diagram

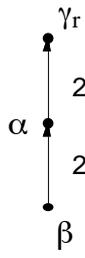

with $\mu_2 = e_2$ , $\beta = e_1 - e_2$ , $\gamma_r = e_1 + e_2$ and $\alpha = e_1$

In the weight decomposition $\Lambda = \Lambda_0 + \lambda\varepsilon$ we have for this algebra

$\varepsilon = (1 , 0)$

Let be now the following cases:

## Case I

• $\langle \Lambda_0 , \mu_2 \circledR = m > 1$ or what is the same (because $\langle \Lambda_0 , \gamma_r \circledR = 0$) $\Lambda_0 = \left(-\frac{m}{2} , \frac{m}{2}\right)$

In this case $\lambda_0 = -1$ , then

$$\Lambda = \left(\frac{-m-2}{2} , \frac{m}{2}\right)$$

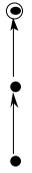

We have only one reduction point, the missing weight in this place will be (Fig. 19)

**Fig. 19**

$$\Lambda - \alpha_0 = \Lambda - \gamma_r = \left(\frac{-m-4}{2} , \frac{m-2}{2}\right)$$

## Case II

• $\langle \Lambda_0 , \mu_2 \circledR = 1$ or, equivalently $\Lambda_0 = \left(-\frac{1}{2} , \frac{1}{2}\right)$

and having in mind that $\Lambda_1 = -1$

$$\Lambda = \left(-1 , \frac{1}{2}\right)$$

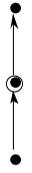

The missing weight will be (Fig. 20)

**Fig. 20**

$$\Lambda - \alpha_0 = \left(-2 , \frac{1}{2}\right)$$

Then the set of unitary representations for the algebra so $(3 , 2)$ $(m \in \mathcal{N}, \lambda \in \leftarrow)$ is the following

a) $\left(\frac{-m-2}{2} , \frac{m}{2}\right)$ $\forall m \geq 2$

Missing weight $\left(\frac{-m-4}{2} , \frac{m-2}{2}\right)$

Extremal vector: $E_{-\beta} E_{-\mu}^2 - (m-1)E_{-\alpha}E_{-\mu} + \frac{m(m-1)}{2}E_{-\gamma_r}$

b) $\left(\frac{-m+2\lambda}{2} , \frac{m}{2}\right)$ $\forall m \geq 2$ $\forall \lambda < -1$

c) $\left(-1 , \frac{1}{2}\right)$

Missing weight $\left(-2 , \frac{1}{2}\right)$

Extremal vector: $E_{-\beta} E_{-\mu} - \frac{1}{2}E_{-\alpha}$

- 38 -

d) $\left(\dfrac{-1+2\lambda}{2}\,,\,\dfrac{1}{2}\right)$     $\forall\lambda<-\dfrac{1}{2}$

e) $\left(\dfrac{-1}{2}\,,\,0\right)$. This highest weight corresponds to the most singular non trivial unitary module $\Lambda_0=(0,0)$.

## X. WAVE EQUATIONS FOR CONFORMAL MULTISPINORS.

For the generators of the fundamental representation of the conformal group, Dirac (DI36) uses the operators:

$$\beta_a=\left(\gamma_1\,,\,\gamma_2\,,\,\gamma_3\,,\,\gamma_4\,,\,\gamma_5\,,\,i\right)$$

$$\gamma_a=\left(\gamma_1\,,\,\gamma_2\,,\,\gamma_3\,,\,\gamma_4\,,\,\gamma_5\,,\,-i\right)$$

where the $\beta's$ and $\gamma's$ satisfy $\beta_a\,\gamma_b+\beta_b\,\gamma_a=2\eta_{ab}$ , $\eta_{ab}=\text{diag}\left(+,\,+,\,+,\,-,\,+,\,-,\right)$ being the metric tensor. On four dimensional conformal space

$$x^2=\eta_{ab}\,x^a x^b=x_1^2+x_2^2+x_3^2-x_4^2+x_5^2-x_6^2=0\quad,\quad x_a=\lambda x_a$$

the conformal group acts linearly. To avoid any dynamics along the rays we fix the degree of homogeneity of all fields by imposing $x^a\,\partial_a\,\psi=n\,\psi$ . The generators for the fundamental representation are then

$$J_{ab}=\frac{1}{i}\left(x_a\,\partial_b-x_b\,\partial_a\right)+\frac{1}{4i}\left(\beta_a\,\gamma_b-\beta_b\,\gamma_a\right)\equiv M_{ab}+S_{ab}$$

The generators of the contragredient representation are obtained by interchanging the $\beta's$ and $\gamma's$ , namely,

$$J_{ab}=\frac{1}{i}\left(x_a\,\partial_b-x_b\,\partial_a\right)+\frac{1}{4i}\left(\gamma_a\,\beta_b-\gamma_b\,\beta_a\right)$$

Both representations have the spin content (1/2 , 0) and (0 , 1/2) respectively. These representations act on some spinor fields of one index $\psi_\alpha$ and $\psi_{\dot\alpha}$ respectively.

Multispinor field transform under the direct product of the fundamental and contragredient representations (2$j$ and 2$k$ times, respectively). They transform under the infinitesimal generators in the following way:

$$\delta\psi_{\alpha_1\,\alpha_2\,\ldots\,\alpha_{2j}}=\varepsilon^{ab}\,\frac{1}{2i}\sum_{i=1}^{2j}\left(\beta_a\,\gamma_b\right)_{\alpha_i}^{\alpha_i'}\,\psi_{\alpha_1\,\ldots\,\alpha_i'\,\ldots\,\alpha'_{2j}}$$

$$\delta\psi_{\dot\alpha_1\,\dot\alpha_2\,\ldots\,\dot\alpha_{2k}}=\varepsilon^{ab}\,\frac{1}{2i}\sum_{i=1}^{2k}\left(\gamma_a\,\beta_b\right)_{\dot\alpha_i}^{\dot\alpha_i'}\,\psi_{\dot\alpha_1\,\ldots\,\dot\alpha_i'\,\ldots\,\dot\alpha_{2k}}$$

where the multispinors are traceless and totally symmetric in the $\alpha_i$ and $\alpha_i$ . They correspond to the representations $(j\,,\,0\,;\,j+1)$ and $(0\,,\,k\,;\,k+1)$ in the Yao (YA67) classification. All these representations are defined on the light cone $\left(x^2=0\right)$ and therefore, using homogeneous coordinates on a 4-dimensional manifold, when restricted to the Poincaré group, they become massless particles of helicity $j$ and $k$, respectively (MA77).

The wave equation will be worked out with the help of the second order Casimir operator, i.e.

$$C_2=\frac{1}{2}\,J_{ab}\,J^{ab}=\frac{1}{2}\,M_{ab}\,M^{ab}+\frac{1}{2}\,S_{ab}\,S^{ab}=\frac{1}{2}\,M^2+MS+\frac{1}{2}\,S^2$$

Each term of the Casimir operator commutes with the others and with the operator

$x^a \partial_a \equiv x \partial$. . Therefore, they must have a common set of eigenfunctions, which are at the same time, the carrier space where the representations acts.

Applying $S^2$ to the highest state of some irreducible representation $(j, 0 ; j+1)$ we get

$$\frac{1}{2} S^2 = 2j(j+1) + j(j+4)$$

and similarly for the representation $(0, k ; k+1)$

$$\frac{1}{2} S^2 = 2k(k+1) + k(k+4)$$

The operator $(x\partial)$ has the eigenvalue $n$, i.e. the homogeneity degree. The operator $\frac{1}{2} M^2$ has therefore eigenvalue $n^2 + 4n$ (recall $x^2 = 0$). Collecting all the eigenvalues of the operators $M^2$, $MS$, $S^2$ and $(x\partial)$, we get for the Casimir operator:

$$C_2 = n^2 + 4n + m + 3j(j+2) = 3(j^2 - 1)$$

and similar expresion for the representation $(0, k ; k+1)$

$$C_2 = n^2 + 4n + m + 3k(k+2) = 3(k^2 - 1)$$

The eigenvalue equation for the operator $MS$ gives

$$\sum_{i=1}^{2j} (\beta x)_i (\gamma \partial)_i \psi = (2jn - m)\psi \quad \text{for} \quad (j, 0 ; j+1)$$

with two cases:

- $\boxed{2jn \neq m}$ . Applying $\prod_i (\gamma \partial)_i$ from the left, we get

$$2j(6 + 2(n-1)) \prod_i (\gamma \partial)_i \psi = (2jn - m) \prod_i (\gamma \partial)_i \psi$$

Therefore $m = -2j(2j+1)$, $n = 2j - 3$ ; $m = -6j$, $n = -1$

- $\boxed{2jn = m}$ . From the Casimir operator, we get

$$n = -(2j+1) , \quad m = -2j(2j+1) ; \quad n = -3 , \quad m = -6j.$$

For $n = -(2j+1)$ we have the field $\psi$ satisfying

$$\sum_{i=1}^{2j} (\beta x)_i (\gamma \partial)_i \psi \equiv Q\psi = 0$$

They correspond to the physical and gauge modes in the Gupta - Bleuler formalism, of electrodynamics as it will be shown.

Their defining equation corresponds to a generalized "Lorentz condition", $\partial^\mu A_\mu = 0$ .

For $n = -1$, we have the field $\chi$ satisfying

$$\sum_{i=1}^{2j} (\beta x)_i (\gamma \partial)_i \chi = 4j\chi$$

It must describe massless particle irreducibly as the field strength $F_{\mu\nu}$ does.

Between the physical and the irreducible mode we impose the condition

$$\chi = \prod_{i=1}^{2j} (\gamma x)_i \psi$$

equivalent to the relation between the field potential and the field strength

$$\left( F_{\mu\nu} = \partial_\mu A_\nu - \partial_\nu A_\mu \right)$$

Now we want an explicit realization for the solutions of the field equations. In the representation $D(j, 0; j+1)$ the lowest state corresponding to the massless physical state can be written as:

$$\psi_p^0 = \frac{1}{x_+^{2j+1}} \underbrace{\begin{pmatrix} 0 \\ 1 \\ 0 \\ 0 \end{pmatrix} \times \ldots \times \begin{pmatrix} 0 \\ 1 \\ 0 \\ 0 \end{pmatrix}}_{2j \text{ terms}}$$

the weight of which is $(j, 0; j+1)$. Using the raising generators of the so(4,2) algebra we can construct all the states where the (indecomposable) representations acts. These states satisfy the same field equation as the lowest state, although they have different weights (HL88)

The lowest state of the gauge mode has the realization

$$\psi_g^0 = \frac{1}{x_+^{2j+2}} \underbrace{\begin{pmatrix} 0 \\ 0 \\ 0 \\ 1 \end{pmatrix} \times \begin{pmatrix} 0 \\ 1 \\ 0 \\ 0 \end{pmatrix} \times \ldots \times \begin{pmatrix} 0 \\ 1 \\ 0 \\ 0 \end{pmatrix}}_{2j \text{ terms}} + \text{ symm. terms}$$

with weight $\left(-j+\frac{1}{2}, -\frac{1}{2}; j+2\right)$. It corresponds to the extremal vector Y obtained by the method outlined before, namely

$$\psi_g^0 = Y\psi_p^0 = \left(E_\beta E_{\mu_2} - 2j E_{\alpha_2}\right)\psi_p^0$$

Therefore it generates an invariant subspace in the representation defined by the physical state $\psi_p^\vee$. If we take the quotient space with respect to this invariant subspace, we are left with the unitary (infinite dimensional) representation corresponding to the massless particle of helicity $j$.

The field $\chi$ has the lowest state

$$\chi^0 = \frac{1}{x_+^{2j+1}} \underbrace{\left(\gamma x\right)\begin{pmatrix} 0 \\ 1 \\ 0 \\ 0 \end{pmatrix} \times \ldots \times \left(\gamma x\right)\begin{pmatrix} 0 \\ 1 \\ 0 \\ 0 \end{pmatrix}}_{2j \text{ terms}}$$

with weight $(j, 0; j+1)$. It generates an irreducible representation because it does not contain extremal vectors. In fact, it can be checked:

$$Y\chi^0 = \left(E_\beta E_{\mu_2} - 2j E_{\alpha_2}\right)\chi^0 = 0$$

In order to complete the Gupta - Bleuler formalism we need also a scalar field $\psi_s$ (corresponding to the longitudinal component of the electromagnetic field) with the following properties:

i) it has the same weight as the corresponding gauge field.

ii) it does not belong to the envelopping algebra (i.e. it can not be obtained by some raising generators applied to $\psi_p^0$).

iii) it is a cyclic state, i.e. the physical lowest state $\psi_p^0$ and the states of the (indecomposable) representation can be obtain from this scalar state. It can be realized by the state

$$\psi_s^0 = \frac{1}{x_+^{2j+2}} \begin{pmatrix} 0 \\ 0 \\ 0 \\ 1 \end{pmatrix} \times \begin{pmatrix} 0 \\ 1 \\ 0 \\ 0 \end{pmatrix} \times \ldots \times \underbrace{\begin{pmatrix} 0 \\ 1 \\ 0 \\ 0 \end{pmatrix}}_{\text{2j terms}} + \text{symm. terms}$$

with weight $\left(-j + \frac{1}{2}, -\frac{1}{2}; j+2\right)$ satisfing $E_{-\alpha_2} \psi_s^0 = \psi_p^0$ , as required. Finally for the scalar field we have $n = -(2j+1)$ , and the following equation holds:

$$Q \psi_s^0 = \psi_q^0 \quad, \quad \text{hence} \quad, \quad Q^2 \psi_s = 0$$

The three fields, scalar, physical and gauge have the same degree of homogeneity and satisfy the same fields equation.

Among the corresponding representation the following leakage takes place:

$$D\left(j - \frac{1}{2}, \frac{1}{2}; j+2\right) \to D(j, 0; j+1) \to D\left(j - \frac{1}{2}, \frac{1}{2}; j+2\right)$$

hence, they constitue a Gupta - Bleuler triplet.

# XI. WAVE EQUATIONS IN DE SITTER SPACE

The coordinates $x_a$, $a = 1, 2, 3, 4, 6$ satisfy

$$x^2 = \eta_{a\beta} x^a x^b = x_1^2 + x_2^2 + x_3^2 - x_4^2 - x_6^2 < 0 \quad, \quad x = \lambda x \quad, \quad \lambda > 0$$

On the four dimensional representation the generators are

$$J_{ab} = -i\left(x_a \partial_b - x_b \partial_a\right) - \frac{i}{4}\left(\beta_a \gamma_b - \beta_b \gamma_a\right)$$

where $\beta_a \gamma_b + \beta_b \gamma_a = 2\eta_{ab}$

We consider the fully symmetric multispinor fields $\psi_{\{\alpha_1 \ldots \alpha_{2j}\}}(x)$ and $\psi_{\{\dot\alpha_1 \ldots \dot\alpha_{2j}\}}$ which transform under de Sitter group in the same way as the conformal group. The maximal compact subgroup SO(3) x SO(2) of SO(3,2) gives us the labels of the representations $\left(E_0, j\right)$ , $E_0$ being the energy of the system and $j$ the angular momentum.

The wave equation is constructed with the help of the second order Casimir operator

$$C_2 = \frac{1}{2} J_{ab} J^{ab} = \frac{1}{2} M^2 + MS + \frac{1}{2} S^2 = E_0\left(E_0 - 3\right) + j\left(j+1\right)$$

with the same notation as in the conformal case; we have for the eigenvalues of the commuting operators

$$\frac{1}{2} M^2 = n^2 + 3n \quad, \quad MS = m \quad, \quad \frac{1}{2} S^2 = 2j\left(j+2\right)$$

for some common eigenstate in the representation $D\left(E, j\right)$.

Notice that we do not work in the light cone $\left(x^2 = 0\right)$ but we suppose the scalar field condition $\left(\partial^2 = 0\right)$ . We have the following cases

i) $D\left(\frac{1}{2}, 0\right) \quad m = 0 \quad, \quad n = -\frac{1}{2} \quad : \quad \partial^2 \psi = 0$

It corresponds to the scalar representation founded by Dirac ("RAC" in Fronsdal's notation)

ii)  $D\left(1,\frac{1}{2}\right)$  $m = n = -\frac{3}{2}$  :  $(\beta x)(\gamma \partial)\,\psi = 0$

$$= -\frac{5}{2}\ ,\ n = -\frac{1}{2}\ :\ (\beta x)(\gamma \partial)\,\chi = 2\chi$$

It corresponds to the spinor representation founded by Dirac ("DI" in Fronsdal's notation)

iii) General case $D(j+1,j)$

$$n = -1\ ,\ m = -4j\ :\ \sum_{i=1}^{2j} (\beta x)_i\, (\gamma \partial)_i\, \chi = 2j\chi$$

$$n = -(2j+1)\ ,\ m = -2j(2j+1)\ :\ \sum_{i=1}^{2j} (\beta x)_i\, (\gamma \partial)_i\, \psi = 0$$

It corresponds to the unitary representation of lowest weight $(E_0, j)$ for massless particles (EV67)

An explicit realization of the solutions for these equations in the general case is given by the lowest state for the physical state (HL90)

$$\psi_p^0 = \frac{1}{x_+^{2j+1}} \begin{pmatrix} 0 \\ 1 \\ 0 \\ 0 \end{pmatrix} \times \ldots \times \begin{pmatrix} 0 \\ 1 \\ 0 \\ 0 \end{pmatrix}$$

$$\underbrace{\qquad\qquad}_{\text{2j terms}}$$

with weight $(j+1,j)$. This lowest state defines an indecomposable representation on the envelopping algebra of so$(3,2)$.

The lowest state of the gauge field corresponding to the invariant subspace generated by the extremal vector, namely, is:

$$\psi_g^0 = \left( E_\beta\, E_\mu^2 + (2j-1)\, E_{\alpha_1}\, E_\mu + (2j-1) j\, E_{\alpha_2} \right) \psi_p^0 =$$

$$= \frac{(2j+1)\,i\,(\beta u\,\gamma_5)}{(2j+1)} \left[ (2j-1) \begin{pmatrix} 1 \\ 0 \\ 0 \\ 0 \end{pmatrix} \times \begin{pmatrix} 0 \\ 1 \\ 0 \\ 0 \end{pmatrix} \times \ldots \times \begin{pmatrix} 0 \\ 1 \\ 0 \\ 0 \end{pmatrix} - \right.$$

$$\underbrace{\qquad\qquad\qquad}_{\text{2j terms}}$$

$$\left. - \begin{pmatrix} 0 \\ 1 \\ 0 \\ 0 \end{pmatrix} \times \begin{pmatrix} 1 \\ 0 \\ 0 \\ 0 \end{pmatrix} \times \ldots \times \begin{pmatrix} 0 \\ 1 \\ 0 \\ 0 \end{pmatrix} - \ldots - \begin{pmatrix} 0 \\ 1 \\ 0 \\ 0 \end{pmatrix} \times \ldots \times \begin{pmatrix} 0 \\ 1 \\ 0 \\ 0 \end{pmatrix} \times \begin{pmatrix} 1 \\ 0 \\ 0 \\ 0 \end{pmatrix} + \text{symm.} \right]$$

$$\underbrace{\qquad\qquad}_{\text{2j terms}} \qquad\qquad \underbrace{\qquad\qquad}_{\text{2j terms}}$$

with weight $(j+2, j-1)$. If we take the quotient space of the indecomposable representation with respect to this invariant subspace we are left with the unitary representation representing massless particles of spin $j$. The scalar state

$$\psi_s^0 = \frac{(2j+1)}{2x_+^{2j+1}} \left[ (2j-1) \underbrace{\begin{pmatrix}0\\0\\1\\0\end{pmatrix} \times \begin{pmatrix}0\\1\\0\\0\end{pmatrix} \times \ldots \times \begin{pmatrix}0\\1\\0\\0\end{pmatrix}}_{2j \text{ terms}} - \right.$$

$$\left. - \underbrace{\begin{pmatrix}0\\0\\0\\1\end{pmatrix} \times \begin{pmatrix}1\\0\\0\\0\end{pmatrix} \times \begin{pmatrix}0\\1\\0\\0\end{pmatrix} \times \ldots \times \begin{pmatrix}0\\1\\0\\0\end{pmatrix}}_{2j \text{ terms}} - \underbrace{\begin{pmatrix}0\\0\\0\\1\end{pmatrix} \times \begin{pmatrix}0\\1\\0\\0\end{pmatrix} \times \ldots \times \begin{pmatrix}0\\1\\0\\0\end{pmatrix} \times \begin{pmatrix}1\\0\\0\\0\end{pmatrix}}_{2j \text{ terms}} \right]$$

is cyclic leaking to the physical state

$$E_{-\alpha_2} \psi_s^0 = -ij\left(4j^2 - 1\right) \psi_p^0 .$$

The three states $\psi_s^{\smile}$, $\psi_p^{\smile}$, $\psi_g^{\smile}$ define a Gupta - Bleuler triplet, whose representations, are leaking in the following way:

$$D(j+2, j-1) \rightarrow D(j+1, j) \rightarrow D(j+2, j-1)$$

## APPENDIX. VERMA MODULES FOR THE CLASSICAL ALGEBRAS

**$A_{n-1}$**

*Generators:*

$H_i = |i\rangle\langle i|$ with $1 \leq i \leq n$; $G_{ij} = |i\rangle\langle j|$ ; $\overline{G}_{ij} = |j\rangle\langle i|$ with $1 \leq i < j \leq n$ where $\langle i|$ denotes the file vector with a 1 at the $i$ position and 0 at the rest of the components, $|i\rangle$ is the corresponding column vector.

*Basis in $\Omega_-$:*

$$\overline{G}_{12}^{g_{12}} \ \overline{G}_{13}^{g_{13}} \ldots \overline{G}_{1n}^{g_{1n}} \ \overline{G}_{23}^{g_{23}} \ldots \overline{G}_{2n}^{g_{2n}} \ldots \overline{G}_{n-1,n}^{g_{n-1,n}} \equiv \overline{X}\left(g_{12} \ldots g_{n-1,n}\right)$$

Representation with highest weight $\Lambda = \left(\Lambda_1, \Lambda_2, \ldots, \Lambda_n\right)$ :

$$\rho\left(H_i\right)\overline{X} = \left(\Lambda_i + \sum_{k=1}^{i-1} g_{ki} - \sum_{k=i+1}^{n} g_{ik}\right)\overline{X}$$

$$\rho\left(\overline{G}_{ij}\right)\overline{X} = \overline{X}\left(g_{ij}+1\right) + \sum_{k=1}^{i-1} g_{ki} \ \overline{X}\left(g_{ki}-1, g_{kj}+1\right)\left(g_{ki}-1, g_{kj}+1\right)$$
$$(j=i+1)$$

$$\rho\left(G_{ij}\right)\overline{X} = \sum_{k=1}^{i-1} g_{kj} \ \overline{X}\left(g_{ki}+1, g_{kj}-1\right) +$$
$$(j=i+1)$$
$$+ g_{ij}\left(\Lambda_i - \Lambda_j + 1 + \sum_{k=j+1}^{n} g_{jk} - \sum_{k=i+1}^{n} g_{ik}\right)\overline{X}\left(g_{ij}-1\right) -$$
$$- \sum_{k=j+1}^{n} g_{ik} \ \overline{X}\left(g_{ik}-1, g_{jk}+1\right)$$

where, for the sake of simplicity we only show the $\overline{X}$ content when some of its coefficients change

*Extremal vectors in $\Omega_-$:*

The vector $Y$ is extremal if it satisfies $\rho(G_{ij})Y = 0$ and its weight $M = (M_1, \ldots, M_n)$ may be obtained from the relation $\rho(H_i)Y = M_i Y$ with

$$\sum_{i=1}^{n} M_i = 0.$$

The extremal vectors with only one generator are of the form

$$Y = \overline{G}_{ij}^{\Lambda_i - \Lambda_j + 1} \quad ; \quad (j = i + 1)$$

with weight : $M = (\Lambda_1, \Lambda_2, \ldots, \Lambda_{i-1}, \Lambda_{i+1} - 1, \Lambda_i + 1, \ldots, \Lambda_n)$

## $B_n$ :

### *Generators :*

$H_i = |2i-1\rangle\langle 2i-1| - |2i\rangle\langle 2i|$ $\qquad \overline{E}_i = |2n+1\rangle\langle 2i-1| - |2i\rangle\langle 2n+1|$

$E_i = |2i-1\rangle\langle 2n+1| - |2n+1\rangle\langle 2i|$ $\qquad \overline{F}_{ij} = |2j\rangle\langle 2i-1| - |2i\rangle\langle 2j-1|$ $\quad (i < j)$

$F_{ij} = |2i-1\rangle\langle 2j| - |2j-1\rangle\langle 2i|$ $\qquad \overline{G}_{ij} = |2j-1\rangle\langle 2i-1| - |2i\rangle\langle 2j|$ $\quad (i < j)$

$G_{ij} = |2i-1\rangle\langle 2j-1| - |2j\rangle\langle 2i|$

### *Basis in $\Omega_-$ :*

$\overline{E}_1^{\,e_1} \ldots \overline{E}_n^{\,e_n}\, \overline{F}_{12}^{\,f_{12}}\, \overline{F}_{13}^{\,f_{13}} \ldots \overline{F}_{1n}^{\,f_{1n}}\, \overline{F}_{23}^{\,f_{23}} \ldots \overline{F}_{n-1,n}^{\,f_{n-1,n}}\, \overline{G}_{12}^{\,g_{12}} \ldots \overline{G}_{n-1,n}^{\,g_{n-1,n}} \equiv \overline{X}(e_1, \ldots, g_{n-1,n})$

Representation with highest weight $\Lambda = (\Lambda_1, \Lambda_2, \ldots, \Lambda_n)$

$$\rho(H_i)\overline{X} = \left[ \Lambda_i - e_i - \sum_{k=i+1}^{n}(f_{ik} + g_{ik}) - \sum_{l=1}^{i-1}(f_{li} - g_{li}) \right]\overline{X}$$

$$\rho(E_n)\overline{X} = -e_n\left[ -\Lambda_n + \frac{e_n - 1}{2} + \sum_{i=1}^{n-1}(e_i - g_{in}) \right]\overline{X}(e_n - 1) +$$

$$+ \frac{e_n(e_n - 1)}{2}\sum_{i=1}^{n-1}e_i\, \overline{X}(e_i - 1, e_n - 2, f_{in} + 1) +$$

$$+ e_n\sum_{i=1}^{n-2}\sum_{j=i+1}^{n-1}e_i e_j\, \overline{X}(e_i - 1, e_j - 1, e_n - 1, f_{ij} + 1) -$$

$$- \sum_{i=2}^{n-1}\sum_{j=1}^{i-1}e_i f_{jn}\, \overline{X}(e_i - 1, f_{ji} + 1, f_{jn} - 1) +$$

$$+ \sum_{i=1}^{n-2}\sum_{j=i+1}^{n-1}e_i f_{jn}\, \overline{X}(e_i - 1, f_{ij} + 1, f_{jn} - 1) +$$

$$+ \sum_{i=2}^{n-1}\sum_{l=1}^{i-1}e_i g_{li}\, \overline{X}(e_i - 1, g_{li} - 1, g_{ln} + 1) -$$

$$- \sum_{j=1}^{n-1}f_{jn}\, \overline{X}(e_j + 1, f_{jn} - 1) + \sum_{i=1}^{n-1}e_i\, \overline{X}(e_i - 1, g_{in} + 1) +$$

$$+ \sum_{j=1}^{n-2} \sum_{k=j+1}^{n-1} f_{jn} e_k \; \overline{X} \left( e_k - 1 \, , f_{jk} + 1 \, , f_{jn} - 1 \right)$$

$$\rho \left( G_{lm} \right) \overline{X} = g_{lm} \left[ \Lambda_l - \Lambda_m - \left( g_{lm} - 1 \right) + \sum_{j=m+1}^{n} \left( g_{mj} - g_{lj} \right) \right] \overline{X} \left( g_{lm} - 1 \right) -$$
$$\hspace{-4em} {}_{(m=l+1)}$$
$$- e_l \; \overline{X} \left( e_l - 1 \, , e_m + 1 \right) + \frac{e_l \left( e_l - 1 \right)}{2} \; \overline{X} \left( e_l - 2 \, , f_{lm} + 1 \right) -$$

$$- \sum_{j=1}^{l-1} f_{jl} \; \overline{X} \left( f_{jl} - 1 \, , f_{jm} + 1 \right) - \sum_{k=m+1}^{n} f_{lk} \; \overline{X} \left( f_{lk} - 1 \, , f_{mk} + 1 \right) -$$

$$- \sum_{j=m+1}^{n} g_{lj} \; \overline{X} \left( g_{lj} - 1 \, , g_{mj} + 1 \right) + \sum_{i=1}^{l-1} g_{im} \; \overline{X} \left( g_{il} + 1 \, , g_{im} - 1 \right)$$

$$\rho \left( \overline{E}_n \right) \overline{X} = \overline{X} \left( e_n + 1 \right) - \sum_{j=1}^{n-1} e_j \; \overline{X} \left( e_j - 1 \, , f_{jn} + 1 \right)$$,

$$\rho \left( \overline{G}_{lm} \right) \overline{X} = - e_m \, \overline{X} \left( e_l + 1 \, , e_m - 1 \right) + \frac{e_m \left( e_m - 1 \right)}{2} \overline{X} \left( e_m - 2 \, , f_{lm} + 1 \right) -$$
$$\hspace{-4em} {}_{(m=l+1)}$$
$$- \sum_{j=1}^{l-1} f_{jm} \; \overline{X} \left( f_{jl} + 1 \, , f_{jm} - 1 \right) - \sum_{k=m+1}^{n} f_{mk} \; \overline{X} \left( f_{lk} + 1 \, , f_{mk} - 1 \right) +$$

$$+ \sum_{i=1}^{l-1} g_{il} \; \overline{X} \left( g_{il} - 1 \, , g_{im} + 1 \right) + \overline{X} \left( g_{lm} + 1 \right)$$

***Extremal vectors in $\Omega_-$ :***

$$\rho \left( E_n \right) Y = \rho \left( G_{lm} \right) Y = 0$$
$$\rho \left( H_i \right) Y = M_i \, Y$$

The solutions with only one generator :

$i$) $\; Y = \overline{E}_n^{\, 2\Lambda_n + 1} \qquad ; \; M = \left( \Lambda_1 \, , \, \dots \, , \, \Lambda_{n-1} \, , \, - \Lambda_n - 1 \right)$

$ii$) $\; Y = \overline{G}_{ij}^{\, \Lambda_i - \Lambda_j + 1} \quad ; \; M = \left( \Lambda_1 \, , \, \dots \, , \, \Lambda_{i-1} \, , \, \Lambda_{i+1} - 1 \, , \, \Lambda_i + 1 \, , \, \Lambda_{i+2} \, , \, \dots \, , \, \Lambda_n \right)$
$\hspace{2em} {}_{(j=i+1)}$

## $C_n$ :

***Generators :***

$$H_i = \begin{pmatrix} |i\rangle\langle i| & 0 \\ 0 & -|i\rangle\langle i| \end{pmatrix}$$

$$E_i = \begin{pmatrix} 0 & |i\rangle\langle i| \\ 0 & 0 \end{pmatrix} \qquad\qquad \overline{E}_i = \begin{pmatrix} 0 & 0 \\ |i\rangle\langle i| & 0 \end{pmatrix} \quad 1 \le i \le n$$

$$G_{ik} = \begin{pmatrix} |i\rangle\langle k| & 0 \\ 0 & -|k\rangle\langle i| \end{pmatrix} \qquad\qquad \overline{G}_{ik} = \begin{pmatrix} |k\rangle\langle i| & 0 \\ 0 & -|i\rangle\langle k| \end{pmatrix}$$

$$F_{ik} = \begin{pmatrix} 0 & |i\rangle\langle k| + |k\rangle\langle i| \\ 0 & 0 \end{pmatrix} \qquad \overline{F}_{ik} = \begin{pmatrix} 0 & 0 \\ |i\rangle\langle k| + |k\rangle\langle i| & 0 \end{pmatrix}$$

***Basis in $\Omega_-$ :***

$$\overline{E}_1^{\,e_1} \ldots \overline{E}_n^{\,e_n} \, \overline{F}_{12}^{\,f_{12}} \, \overline{F}_{13}^{\,f_{13}} \ldots \overline{F}_{1n}^{\,f_{1n}} \, \overline{F}_{23}^{\,f_{23}} \ldots \overline{F}_{n-1,n}^{\,f_{n-1,n}} \, \overline{G}_{12}^{\,g_{12}} \ldots \overline{G}_{n-1,n}^{\,g_{n-1,n}} \equiv \overline{X}\left(e_1, \ldots, g_{n-1,n}\right)$$

Representation with highest weight $\Lambda = \left(\Lambda_1, \Lambda_2, \ldots, \Lambda_n\right)$ :

$$\rho\left(H_i\right)\overline{X} = \left[\Lambda_i - 2e_i - \sum_{j=1}^{i-1}(f_{ji} - g_{ji}) - \sum_{j=i+1}^{n}(f_{ij} + g_{ij})\right]\overline{X}$$

$$\rho\left(E_n\right)\overline{X} = -e_n\left[-\Lambda_n + (e_n - 1) + \sum_{j=1}^{n-1}(f_{jn} - g_{jn})\right]\overline{X}\left(e_n - 1\right) -$$

$$- \sum_{i=1}^{n-1} f_{in}(f_{in} - 1)\overline{X}\left(e_i + 1, f_{in} - 2\right) -$$

$$- \sum_{i=1}^{n-2}\sum_{k=i+1}^{n-1} f_{in} f_{kn} \overline{X}\left(f_{ik} + 1, f_{in} - 1, f_{kn} - 1\right) +$$

$$+ \sum_{i=2}^{n-1}\sum_{k=1}^{i-1} f_{in} g_{ki}\overline{X}\left(f_{in} - 1, g_{ki} - 1, g_{kn} + 1\right) +$$

$$+ \sum_{i=1}^{n-1} f_{in} \, \overline{X}\left(f_{in} - 1, g_{in} + 1\right)$$

$$\rho\left(G_{ij}\right)\overline{X} = -e_i \, \overline{X}\left(e_i - 1, f_{ij} + 1\right) - 2f_{ij} \overline{X}\left(e_j + 1, f_{ij} - 1\right) -$$
$$\scriptstyle (i = i+1)$$

$$- \sum_{m=j+1}^{n} f_{im}\overline{X}\left(f_{im} - 1, f_{jm} + 1\right) - \sum_{l=1}^{i-1} f_{li}\overline{X}\left(f_{li} - 1, f_{lj} + 1\right) -$$

$$- \sum_{m=j+1}^{n} g_{im}\overline{X}\left(g_{im} - 1, g_{jm} + 1\right) + \sum_{l=1}^{i-1} g_{lj}\overline{X}\left(g_{li} + 1, g_{lj} - 1\right) +$$

$$+ g_{ij}\left[\Lambda_i - \Lambda_j - (g_{ij} - 1) + \sum_{l=j+1}^{n}(g_{jl} - g_{il})\right]\overline{X}\left(g_{ij} - 1\right)$$

$$\rho\left(\overline{E}_n\right)\overline{X} = \overline{X}\left(e_n + 1\right)$$

$$\rho\left(\overline{G}_{ij}\right)\overline{X} = -e_j \, \overline{X}\left(e_j - 1, f_{ij} + 1\right) - \sum_{l=1}^{i-1} f_{lj} \, \overline{X}\left(f_{li} + 1, f_{lj} - 1\right) -$$
$$\scriptstyle (i = i+1)$$

$$- 2f_{ij} \, \overline{X}\left(e_i + 1, f_{ij} - 1\right) - \sum_{m=j+1}^{n} f_{jm} \, \overline{X}\left(f_{im} + 1, f_{jm} - 1\right) +$$

$$+ \overline{X}\left(g_{ij} + 1\right) + \sum_{k=1}^{i-1} g_{ki} \, \overline{X}\left(g_{ki} - 1, g_{kj} + 1\right) -$$

$$- \sum_{l=j+1}^{n} g_{jl} \ \overline{X} \ \left( g_{il} + 1 \ , \ g_{jl} - 1 \right)$$

**Extremal vectors in $\boldsymbol{\Omega_-}$ :**

$$\rho\left(E_n\right) Y \ = \ \rho\left(G_{ij}\right) Y \ = \ 0$$
$$\rho\left(H_i\right) Y \ = \ M_i \, Y$$

The solutions with only one generator:

$i$) $Y \ = \ \overline{E}_n^{\,\Lambda_n+1}$ ; $M = \left(\Lambda_1 \ , \ \Lambda_2 \dots , \ - \Lambda_n - 2\right)$

$ii$) $Y \ = \ \overline{G}_{i,i+1}^{\,\Lambda_i - \Lambda_{i+1} + 1}$ ; $M = \left(\Lambda_1 \ , \ \Lambda_3 \dots , \ \Lambda_{i-1} \ , \ \Lambda_{i+1} - 1 \ , \ \Lambda_i + 1 \ , \ \Lambda_{i+2} \ , \ \dots \ , \ \Lambda_n\right)$

**Dn :**

**Generators :**

$$H_i \ = \ |\, 2i-1 \,\rangle \langle\, 2i-1 \,| - |\, 2i \,\rangle \langle\, 2i \,| \ \ ; \ \ 1 \le i \le n$$
$$F_{ij} \ = \ |\, 2i-1 \,\rangle \langle\, 2j \,| - |\, 2j-1 \,\rangle \langle\, 2i \,| \quad \overline{F}_{ij} \ = \ |\, 2j \,\rangle \langle\, 2i-1 \,| - |\, 2i \,\rangle \langle\, 2j-1 \,|$$
$$G_{ij} \ = \ |\, 2i-1 \,\rangle \langle\, 2j-1 \,| - |\, 2j \,\rangle \langle\, 2i \,| \quad \overline{G}_{ij} \ = \ |\, 2j-1 \,\rangle \langle\, 2i-1 \,| - |\, 2i \,\rangle \langle\, 2j \,|$$

$1 \le i < j \le n$

**Basis in $\boldsymbol{\Omega_-}$ :**

$$\overline{F}_{12}^{\,f_{12}} \ \overline{F}_{13}^{\,f_{13}} \ \dots \ \overline{F}_{1n}^{\,f_{1n}} \ \overline{F}_{23}^{\,f_{23}} \ \dots \ \overline{F}_{n-1,n}^{\,f_{n-1,n}} \ \overline{G}_{12}^{\,g_{12}} \ \dots \ \overline{G}_{n-1,n}^{\,g_{n-1,n}} \ \equiv \ \overline{X}\left(f_{12} \ , \ \dots \ , \ g_{n-1,n}\right)$$

Representation with highest weight $\Lambda = \left(\Lambda_1 \ , \ \Lambda_2 \ , \ \dots \ , \ \Lambda_n\right)$

$$\rho\left(H_i\right) \overline{X} \ = \ \left[ \Lambda_i - \sum_{k=i+1}^{n} \left(f_{ik} + g_{ik}\right) - \sum_{l=1}^{i-1} \left(f_{li} - g_{li}\right) \right] \overline{X}$$

$$\rho\left(F_{m,n}\right) \overline{X} \ = \ \sum_{i=1}^{m-1} f_{in} \ \overline{X} \left( f_{in} - 1 \ , \ g_{im} + 1 \right) +$$
$$(n=m+1)$$
$$+ \ \sum_{i=2}^{m-1} \sum_{k=1}^{i-1} f_{in} \, g_{ki} \left(g_{ki} - 1\right) \overline{X}\left(f_{in} - 1 \ , \ g_{ki} - 1\right) +$$
$$+ \ \sum_{i=2}^{m-1} \sum_{n=1}^{i-1} f_{in} \, g_{ki} \, \overline{X}\left(f_{in} - 1 \ , \ g_{ki} - 1 \ , \ g_{km} + 1\right) -$$
$$- \ \sum_{i=1}^{m-2} f_{im} f_{m-1,n} \ \overline{X}\left(f_{i,m-1} + 1 \ , \ f_{im} - 1 \ , \ f_{m-1,n} - 1\right) -$$
$$- \ \sum_{i=1}^{m-1} f_{im} \, \overline{X}\left(f_{im} - 1 \ , \ g_{in} + 1\right) -$$

$$- \sum_{i=2}^{m-1} \sum_{k=1}^{i-1} g_{im} \, g_{ki} \, \overline{X} \left( f_{im} - 1 \, , \, g_{ki} - 1 \, , \, g_{kn} + 1 \right) -$$

$$- f_{mn} \left[ -\Lambda_m - \Lambda_n + (f_{mn} - 1) - \sum_{k=1}^{m-1} \left( g_{km} + g_{kn} \right) \right] \overline{X} \left( f_{mn} - 1 \right) -$$

$$- \sum_{i=1}^{m-1} f_{im} f_{mn} \overline{X} \left( f_{mn} - 1 \right) - \sum_{i=1}^{m-1} f_{in} f_{mn} \overline{X} \left( f_{mn} - 1 \right)$$

$$\rho \left( G_{lm} \right) \overline{X} = g_{lm} \left[ \Lambda_l - \Lambda_m - (g_{lm} - 1) + \sum_{j=m+1}^{n} \left( g_{mj} - g_{lj} \right) \right] \overline{X} \left( g_{lm} - 1 \right) -$$
$$\scriptstyle (m=l+1)$$

$$- \sum_{j=1}^{l-1} f_{jl} \, \overline{X} \left( f_{jl} - 1 \, , \, f_{jm} + 1 \right) - \sum_{k=m+1}^{n} f_{lk} \, \overline{X} \left( f_{lk} - 1 \, , \, f_{mk} + 1 \right) -$$

$$- \sum_{j=m+1}^{n} g_{lj} \, \overline{X} \left( g_{lj} - 1 \, , \, g_{mj} + 1 \right) + \sum_{i=1}^{l-1} g_{im} \, \overline{X} \left( g_{il} + 1 \, , \, g_{im} - 1 \right)$$

$$\rho \left( \overline{F}_{mn} \right) \overline{X} = \overline{X} \left( f_{mn} + 1 \right)$$
$$\scriptstyle (n=m+1)$$

$$\rho \left( \overline{G}_{lm} \right) \overline{X} = - \sum_{j=1}^{l-1} f_{jm} \, \overline{X} \left( f_{jl} + 1 \, , \, f_{jm} - 1 \right) -$$
$$\scriptstyle (m=l+1)$$

$$- \sum_{k=m+1}^{n} f_{mk} \, \overline{X} \left( f_{lk} + 1 \, , \, f_{mk} - 1 \right) +$$

$$+ \sum_{i=1}^{l-1} g_{il} \, \overline{X} \left( g_{il} - 1 \, , \, g_{im} + 1 \right) + \overline{X} \left( g_{lm} + 1 \right)$$

### Extremal vectors in $\Omega_-$ :

$$\rho \left( F_{mn} \right) Y = \rho \left( G_{lm} \right) Y = 0$$
$$\rho \left( H_i \right) Y = M_i Y$$

The solutions with only one generator :

$i$)  $Y = \overline{F}_{n-1, n}^{\Lambda_{n-1} + \Lambda_n + 1}$   ;  $M = \left( \Lambda_1 \, , \, \Lambda_2 \, , \, \dots \, , \, -\Lambda_n - 1 \, , \, -\Lambda_{n-1} - 1 \right)$

$ii$)  $Y = \overline{G}_{m-1, m}^{\Lambda_{m-1} - \Lambda_m + 1}$   ;  $M = \left( \Lambda_1 \, , \, \Lambda_2 \, , \, \dots \, , \, \Lambda_{m-2} \, , \, \Lambda_m - 1 \, , \, \Lambda_{m-1} + 1 \, , \, \dots \, , \, \Lambda_n \right)$

A complete method to calculate all the extremal vectors in Verma modules can be found in LG84  GEL90 and DO90.

**Acknowledgments:**
The authors want to express their gratitude to the Vicerrectorado de Investigación de la Universidad de Oviedo for financial support. This work belongs also to the common project between the Universities of Oviedo (Spain) and Clausthal (Germany) sponsored by the Volkswagen Foundation.

XXXI , 121-126.